\documentclass[journal]{IEEEtran}

\usepackage{times,url,comment}
\usepackage{amsmath,amssymb,amsfonts,bbm}
\usepackage[tight,footnotesize]{subfigure}
\usepackage{epsfig,epsf}
\usepackage{array}
\usepackage{longtable}
\usepackage{multirow}
\usepackage{graphicx}
\usepackage{mdwmath}
\usepackage{mdwtab}
\usepackage{eqparbox}
\usepackage{booktabs}
\usepackage{ctable}
\usepackage{eso-pic}
\usepackage{breqn}
\usepackage{float}
\usepackage{longtable}
\usepackage{enumerate}
\usepackage{textcomp}
\usepackage{hyperref}
\usepackage{doi}
\usepackage{fancyhdr}
\usepackage{comment}
\usepackage{algorithm, algcompatible}
\algnewcommand\algorithmicinput{\textbf{Input:}}
\algnewcommand\INPUT{\item[\algorithmicinput]}
\algrenewcommand{\algorithmiccomment}[1]{\hskip3em$\triangleleft$ #1}
\usepackage{mdframed}

\ifCLASSINFOpdf
\else
\fi

\fancypagestyle{firstpage}
{
    \fancyhead[L]{{\small{© 2022 IEEE.  Personal use of this material is permitted.  Permission from IEEE must be obtained for all other uses, in any current or future media, including reprinting/republishing this material for advertising or promotional purposes, creating new collective works, for resale or redistribution to servers or lists, or reuse of any copyrighted component of this work in other works.  \doi{10.1109/JIOT.2022.3196154}} }}   
}

\begin{document}
\title{Graph Signal Reconstruction Techniques for IoT Air Pollution Monitoring Platforms}
\author{%
Pau Ferrer-Cid,
Jose M. Barcelo-Ordinas,
Jorge Garcia-Vidal
\thanks{Pau Ferrer-Cid (pau.ferrer.cid@upc.edu), Jose M. Barcelo-Ordinas (jose.maria.barcelo@upc.edu) and Jorge Garcia-Vidal (jorge.garcia@upc.edu) are with the Universitat Politecnica de Catalunya, Barcelona, Spain.}
\thanks{This work is supported by the National Spanish funding PID2019-107910RB-I00, by regional project 2017SGR-990, and with the support of Secretaria d’Universitats i Recerca de la Generalitat de Catalunya i del Fons Social Europeu.}
}


\maketitle
\begin{abstract}
Air pollution monitoring platforms play a very important role in preventing and mitigating the effects of pollution. Recent advances in the field of graph signal processing have made it possible to describe and analyze air pollution monitoring networks using graphs. One of the main applications is the reconstruction of the measured signal in a graph using a subset of sensors. Reconstructing the signal using information from neighboring sensors is a key technique for maintaining network data quality, with examples including filling in missing data with correlated neighboring nodes, creating virtual sensors, or correcting a drifting sensor with neighboring sensors that are more accurate. This paper proposes a signal reconstruction framework for air pollution monitoring data where a graph signal reconstruction model is superimposed on a graph learned from the data. Different graph signal reconstruction methods are compared on actual air pollution data sets measuring O$_3$, NO$_2$, and PM$_{10}$. The ability of the methods to reconstruct the signal of a pollutant is shown, as well as the computational cost of this reconstruction. The results indicate the superiority of methods based on kernel-based graph signal reconstruction, as well as the difficulties of the methods to scale in an air pollution monitoring network with a large number of low-cost sensors. However, we show that the scalability of the framework can be improved with simple methods, such as partitioning the network using a clustering algorithm.
\end{abstract}


\thispagestyle{firstpage}
\section{Introduction}
\label{Sec:Intro}
\IEEEPARstart{O}{U}{T}{D}{O}{O}{R} air pollution originates from natural and anthropogenic sources. Air pollution produces adverse health consequences, such as respiratory, neuronal, and cardiovascular problems, among others, as a result of short or long-term exposure. For all these reasons, in recent years much effort has been made to obtain pollution measures for their study and the application of prevention and containment measures. To increase the number of measurement points, the use of low-cost sensors, mounted on Internet of Things (IoT) nodes, has been proposed in conjunction with the values provided by reference stations, forming IoT platforms. Reference stations give much more precise values, but at a much higher price. On the other hand, the greatest difficulty in a deployment with low-cost sensors is the quality of the data it produces. To improve the quality of the low-cost sensors, it is necessary to calibrate \cite{maag2018survey,barcelo2019self} these sensors in the field using supervised machine learning algorithms \cite{esposito2017computational,zimmerman2018machine,ferrer2019comparative,spinelle2015field}. Today, as the calibration of sensors seems to be feasible \cite{williams2019low,ripoll2019testing}, the study of sensor networks for pollution monitoring poses new challenges. Once the sensors are deployed, they suffer from drifts and aging \cite{de2020robustness,miskell2019reliable}, so that the estimates made with the machine learning mechanisms become inaccurate. In addition, both low-cost sensors and reference stations lose data \cite{ferrer2020graphsensing}, which makes the supervised machine learning mechanisms difficult to estimate values.

The study of heterogeneous networks, which contain both reference stations and low-cost sensors, using graphs provides great flexibility, as the graphs can encode complex networks, as is the case when there is a reference station and a sensor in the same geographical location, a situation in which the geostatistical Kriging technique would fail. There is a wide variety of techniques for the estimation of the graph topology based on different methods such as statistical-based methods or graph signal processing (GSP) methods \cite{dong2019learning,mateos2019connecting}.

There are a large number of direct applications of generating a graph with reference stations and low-cost sensors and overlaying a signal reconstruction mechanism in which several of the network nodes participate. The principle behind this technique is to obtain the relationships between the network sensors by means of a graph learned from the data, resulting in a smooth structure with respect to the measured data. Then, the data measured by the different sensor neighborhoods can be used by signal reconstruction methods to obtain estimates and maintain the quality of the network data. An example is the imputation of missing values. It is known that sensors are calibrated with supervised methods using arrays of sensors \cite{esposito2017computational,zimmerman2018machine,ferrer2019comparative,spinelle2015field}. If one of the array measurements is missing, the concentration of the pollutant cannot be estimated and a gap in the measurements occurs \cite{ferrer2020graphsensing,fung2020input,zaidan2019mutual}. This measurement can be estimated using neighbors whose measures are highly correlated with the data from the sensor that has the missing value. Other examples where signal reconstruction methods can be applied overlaid on a graph constructed from the data are sensor drifts \cite{ferrer2019comparative,de2020robustness,barcelo2019distributed}, the creation of virtual sensors \cite{matusowsky2020data} and the creation of proxies \cite{fung2020input,zaidan2019mutual,zaidan2020intelligent}.
Virtual sensors are nodes in which it is difficult to deploy a physical sensor, and in which the value of the pollutant is estimated from values in the vicinity \cite{liu2009virtual}. Thus, graph signal reconstruction can also help maintain network data quality by obtaining virtual sensor estimates when any of the nodes are under maintenance or have been relocated. 
Signal reconstruction is therefore a key technique to benefit from similar sensors, including reference stations, which can correct data for multiple applications.

In this paper, we propose a signal reconstruction framework to maintain air pollution sensor network data quality. Therefore, we study different methods for the reconstruction of the graph signal for several pollutants, studying their reconstruction capabilities when all the nodes have samples, and when various nodes are estimated at the same time. In this way, we demonstrate their ability to estimate air pollution and their scalability for use in IoT air pollution monitoring networks. More precisely, in this study we:

\begin{itemize}
\item propose a signal reconstruction framework based on graph signal reconstruction superimposed on a graph learned from the data. Thus, we compare the performance of three different graph signal reconstruction models applied to a set of Spanish reference stations that measure O$_3$, NO$_2$ and PM$_{10}$ which could have missing values or be virtual sensors,
\item analyze the scalability issues of the techniques for air pollution monitoring networks, and propose an application of clustering techniques to reduce these problems,
\item show a low-cost sensor drift compensation example in a heterogeneous IoT air pollution platform with reference stations and low-cost sensors by using the proposed signal reconstruction framework.
\end{itemize}

The paper is organized as follows: section \ref{Sec:RW} covers the related work. Section \ref{Sec:graph_sensing} describes the graph learning and signal reconstruction techniques used, section \ref{Sec:DS} describes the different Spanish reference station data sets employed. Section \ref{Sec:Res} shows the results for a set of experiments as well as the scalability study. Lastly, section \ref{Sec:Conc} concludes the paper.

\section{Related Work}
\label{Sec:RW}



{\it In-situ calibration and current monitoring networks}:
the use and calibration of low-cost sensors for air pollution monitoring have been intensively studied during the last years \cite{maag2018survey,barcelo2019self}. The calibration process is carried out in uncontrolled environments, which means \cite{barcelo2019self} that the sensor is calibrated in-situ in the field of deployment under real environmental conditions. Machine learning has been a key technique for sensor in-situ calibration, where studies include multiple linear regression  \cite{spinelle2015field,ferrer2019comparative}, random forest \cite{zimmerman2018machine}, support vector regression \cite{ferrer2020multi,esposito2017computational}, and artificial neural  networks \cite{esposito2017computational}. Once the sensors have been calibrated, they are deployed forming a network, and the nodes report the estimated data using a machine learning algorithm employed in the in-situ calibration phase. Some authors also make spatial estimates or create air pollution maps from this data. The most common form of spatial estimation of pollution is Kriging, a geostatistical method that extrapolates the known contamination values in some sources to other geographical locations forming a continuous surface \cite{schneider2017mapping,barcelo2019distributed,shukla2020mapping}. For example, Schneider \textit{et al.} \cite{schneider2017mapping} make use of Kriging to merge values from sensors and reference stations improving this way the estimation, and Barcelo-Ordinas \textit{et al.} \cite{barcelo2019distributed} use Kriging estimates to correct average concentrations based on reference stations measures.

{\it Graph signal processing in air pollution monitoring networks}:
the study of these networks has also been carried out from a discrete point of view, modeling these networks as graphs. In \cite{song2008comparative} it is shown the use of Gaussian Markov random fields (GMRF) for the approximation of Gaussian fields and the extrapolation of environmental measurements. In addition, given the rise of the graph signal processing (GSP) field, which allows the use of classical signal processing techniques on defined signals in irregular structures such as graphs, many kinds of networks are modeled using graphs \cite{mateos2019connecting,ribeiro2018graph}. The Laplacian matrix is the object of study of the GSP and is required to correctly describe the relationship between the network nodes, so there is a lot of literature available regarding the learning of the Laplacian matrix given a set of observations \cite{mateos2019connecting}. In the specific case of environmental monitoring networks, results have already been shown from the use of GSP on such networks. As examples, Jablonski \cite{jablonski2017graph} clusters tropospheric ozone measures in Poland using a GSP framework, and Ferrer-Cid \textit{et al.} \cite{ferrer2020graphsensing} compare three graph inferring methods for building a graph. Graph signal reconstruction techniques have also been used to estimate rainfall and temperature measurements \cite{ribeiro2018graph, romero2016kernel}.

{\it Limitations in current heterogeneous air pollution monitoring networks:}
in-situ calibration, where the signal from an individual sensor is estimated, has several limitations. One is its inability to estimate values at those moments in time when the sensor fails and is unable to give a reading \cite{ferrer2020graphsensing,fung2020input,zaidan2019mutual}. Another is that as the sensor ages or the environmental conditions change from the time the calibration process was performed, sensors drift or age and the estimated data loses quality \cite{barcelo2019distributed,mijling2018field,de2020robustness,miskell2019reliable}. Reference station data sets also report missing data \cite{ferrer2020graphsensing}. The deterioration of the data quality of these sensors can be mitigated if in the estimation phase the signal of a problematic node is reconstructed using other sensors, which are potentially non-problematic, that were highly correlated during the graph learning stage. Thus, graph signal reconstruction for air pollution sensor networks is key to harnessing network information and maintaining data quality, where unobserved nodes (e.g., sensors with missing samples, drifting sensors, etc.) are reconstructed from a subset of observed nodes in the graph, also including places where there are no physical sensors (virtual sensors) \cite{matusowsky2020data,fung2020input}.

{\it Graph signal reconstruction techniques:}
there are methods from the field of semi-supervised learning such as the Laplacian interpolation (Laplacian regularization from Belkin \textit{et al.} \cite{belkin2004regularization}) that use the Laplacian of the graph to extrapolate the observed values to those that are not, maximizing the smoothness of the resulting signal. Other techniques, based on signal processing, use the Fourier transform assuming certain properties of the signal to recover the full signal given some measurements \cite{stankovic2019tutorial}. Some works address the question from the point of view of the kernel methods, kernelizing the signal reconstruction and obtaining the kernel ridge regression (KRR) \cite{romero2016kernel}. Matrix completion methods are transductive methods that complete the missing entries of the data matrix assuming that it has a lower rank representation; e.g. 
 the kernelized probabilistic matrix factorization (KPMF) \cite{zhou2012kernelized} that uses graph kernels to factor the matrix. In addition, another recent field that has benefited from graph signal processing tools is graph neural networks (GNN).
 Among the networks developed we find ChebyNet \cite{defferrard2016convolutional}, which approximates convolution filtering through a Chebyshev polynomial filter, or the inductive graph neural network Kriging (IGNNK) \cite{wu2020inductive}, a 2-layer diffusion convolution neural network that allows reconstructing any of the nodes through a subnetwork selection scheme during the training. But the two last techniques need to be further developed, since matrix completion methods do not naturally fit into the problem of signal reconstruction, as for each new sample it needs to be re-trained. Besides, there not exist much literature about graph neural networks architectures for graph signal reconstruction in cases where data is limited. We restrict the comparison of reconstruction methods to linear models, since non-convex models usually require more data, which are often unavailable in low-cost sensor deployment environments. Moreover, non-convex models are more difficult to adapt to environments where any subset of nodes can have missing data.

{\it Graph signal reconstruction in air pollution monitoring networks:} recently, research has been carried out by applying graph signal reconstruction to PM2.5 data sets \cite{qiu2017time,wang2020pm2}. The matrix completion problem has also been tackled using variational graph autoencoders (VGAE) applied to NO$_2$ and PM data sets \cite{do2020graph}. Graph convolutional recurrent neural networks have also been used for PM data \cite{le2021spatiotemporal}. However, as mentioned above, the matrix completion approach to signal reconstruction, or the need for graph neural networks for large amounts of training data, can make it difficult to use these models in sensor network deployments. In this paper, we opt for the use of linear graph signal reconstruction models on a graph learned from the data, given their low data and computational requirements. This allows extending the use of these techniques to both sensor deployment environments where the data may be scarce, and to environments where any node could fail and the signal reconstruction model would need to be recalculated. Furthermore, in most cases graph signal reconstruction models for air pollution are applied to graphs created from the geodesic distances of the nodes \cite{qiu2017time,do2020graph,le2021spatiotemporal}. However, given the effectiveness of graphs learned from data in air pollution sensor networks \cite{ferrer2020graphsensing}, we chose to learn the graph from the data in our signal reconstruction framework.

{\it Our work:} we propose a framework composed of graph signal reconstruction superimposed on a graph learned from the data to maintain the data quality of a sensor network. Since the data quality maintenance is proportional to the quality of the signal reconstruction, we compare the accuracy of three graph signal reconstruction methods to reconstruct air pollution monitoring signals with missing values for O$_3$, NO$_2$ and PM$_{10}$; Laplacian interpolation, GSP low-pass reconstruction and kernel-based graph signal reconstruction. We propose a cluster-based methodology to split the problem of graph learning and signal reconstruction in order to overcome scalability issues for large air pollution sensor networks. And finally, we show a case study on data quality maintenance where signal reconstruction is used to correct a drifting low-cost sensor.

\begin{table}[!ht]
\centering
\caption{Notation summary.}
\label{tab:notation}
\resizebox{0.9\columnwidth}{!}{%
\begin{tabular}{@{}ccc@{}}
\toprule
\textbf{SYMBOL} & \textbf{} & \textbf{MEANING} \\  \midrule
$N$ $\mid$ $P$ &  & Number of nodes $\mid$ Number of observations \\
$G$ $\mid$ $\mathcal{V}$ $\mid$ $\mathcal{E}$ &  & Graph $\mid$ Set of nodes $\mid$ Set of edges \\
\bf{A} $\mid$ \bf{W} &  & Graph adjacency matrix $\mid$ Graph weight matrix \\
\bf{D} $\mid$ \bf{L} &  & Graph degree matrix $\mid$ Graph Laplacian matrix \\
x$_i$ $\mid$ N($i$) &  & $i$th vertex $\mid$ $i$th vertex neighborhood\\
${\bf\Sigma}$ $\mid$ ${\bf\Theta}$ &  & Covariance matrix $\mid$ Precision matrix \\
tr($\cdot$) $\mid$ det($\cdot$) &  & Trace of a matrix $\mid$ Determinant of a matrix \\
$\|\cdot\|_1$ $\mid$ $\|\cdot\|_F$ &  & l$_1$-norm of a matrix $\mid$ Frobenius norm of a matrix \\
$\alpha$ $\mid$ $\beta$ &  & Graph learning hyperparameters \\
$\bf{U}$ $\mid$ $\bf{K}$ & & Laplacian eigenvector matrix $\mid$ Graph kernel matrix \\
$\mathcal{M}$ $\mid$  $\mathcal{U} $&  & Set of observed $\mid$ unobserved or missing nodes\\
M $\mid$ U & & Number of observed nodes $\mid$ unobserved nodes \\
$\left | \mathcal{M} \right |$ $\mid$ $\bf{U}^{\dagger}$ & & Cardinality of a set $\mid$ Matrix pseudoinverse \\
\bf{0} $\mid$ \bf{1} & & Vector of zeros $\mid$ Vector of ones \\ \bottomrule
\end{tabular}%
}
\end{table}


\section{Graph signal reconstruction framework}
\label{Sec:graph_sensing}
\begin{figure*}[!htb]
    \centering
    \includegraphics[width=0.75\textwidth]{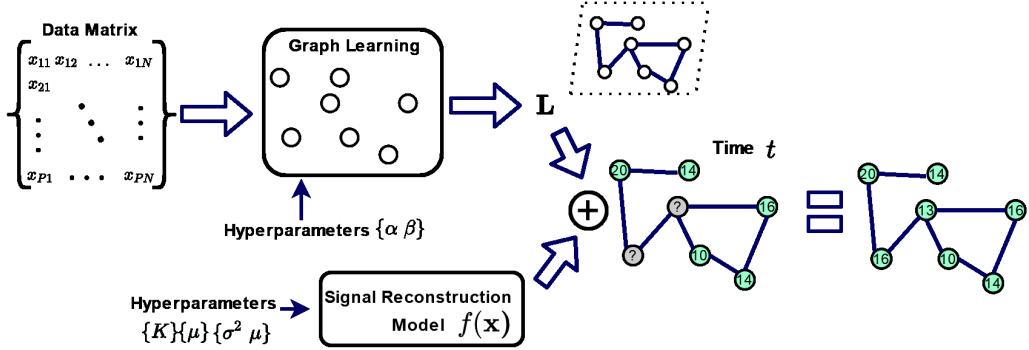}
    \caption{Signal reconstruction framework. Given a set of graph learning and signal reconstruction hyperparameters, the Laplacian can be learned from the data, and network measures can be reconstructed using both the graph and a signal reconstruction model.}
    \label{fig:pipeline}
\end{figure*}

This section presents the framework to learn the graph topology given a set of observations and introduces the signal reconstruction techniques used in the study. Bold uppercase letters denote matrices, bold lowercase letters denote vectors and lowercase letters denote scalars. Table \ref{tab:notation} summarizes the notation used throughout the paper. Figure \ref{fig:pipeline} represents the proposed signal reconstruction framework where a graph signal reconstruction model is superimposed on a graph learned from the sensor network data. The following sections describe the framework's two main pieces; graph learning and signal reconstruction.

\subsection{Graph learning}
\label{subsec:graph_learning}


The graph learning task consists in learning the graph triplet $G$=$\{\mathcal{V},\mathcal{E},\bf{W}\}$ given a set of observations $\bf{X}{\in} \mathbb{R}^{P \times N}$, where $P$ corresponds to the number of observations and $N$ corresponds to the number of nodes. $\mathcal{V}$ is a prefixed set of nodes, $\mathcal{E}$=$\{e_{ij}{:} {v_i, v_j}\in \mathcal{V}\}$ are the learned edges between nodes and $\bf{W}$ is the weight matrix whose entries encode the edges weights. Furthermore, the Laplacian matrix $\bf{L}$ also defines the connectivity of the resulting graph. The Laplacian matrix is defined as $\bf{L}$= $\bf{D}$-$\bf{W}$, where $\bf{D}$ is the diagonal degree matrix with values $D_{mm}$= $\sum_n W_{mn}$. For an undirected graph (our case), the Laplacian matrix is symmetric, $\bf{L}$= ${\bf{L}}^T$. This graph topology can then be used to flow information between similar sensors using signal reconstruction techniques. The nodes that make up the graph are reference stations that provide accurate data, and low-cost sensor array nodes, which provide the estimation of values already calibrated in-situ with supervised machine learning methods.

In recent studies, several approaches have been studied to obtain the weight matrix $\bf{W}$ or the Laplacian matrix $\bf{L}$, from statistical approaches, such as the graphical lasso algorithm, to graph signal processing approaches with distance-based and smoothness-based methods. As most of the reconstruction methods rely on the signal smoothness assumption -vertices connected with large weights tend to have similar values \cite{ribeiro2018graph,mateos2019connecting}- the smoothness method shown by Dong \textit{et al.} \cite{dong2016learning} is used. Dong \textit{et al.} formulate a jointly non-convex problem to find a filtered version $\bf{Y}$ of the observations $\bf{X}$ and the Laplacian matrix $\bf{L}$:
\begin{equation}
\label{eq:dong}
\begin{aligned}
& \underset{\bf{L},\bf{Y}}{\text{min}}
& & \underbrace{\| {\bf{X}} - {\bf{Y}} \|_F^2 }_{data \hspace{0.5em} fidelity} + \underbrace{\underbrace{ \alpha \; tr({\bf{Y}}^T {\bf{L}} {\bf{Y}})}_{smoothness} + \beta \|{\bf{L}}\|_F^2}_{sparsity} \\
& \text{s.t.} & &  tr({\bf L}) = n, \\
& & &  L_{ij} = L_{ji} \leq 0, \;\;\; \;\;\; i\neq j, \\
& & &  {\bf L} \cdot {\bf 1} = {\bf 0}.
\end{aligned}
\end{equation}
This problem is solved by alternating the minimization on $\bf{Y}$ and $\bf{L}$. Two scalar hyperparameters need to be provided to the algorithm; $\alpha$ and $\beta$. Hyperparameter $\alpha$ controls the smoothness of the solution, $\bf{L}$ smooth with respect to $\bf{Y}$, while hyperparameter $\beta$ penalizes the Frobenius norm $||\cdot||_F^2$ of the Laplacian matrix, so controlling the number of non-zero entries of the matrix. As the authors mention, the computational complexity scales quadratically with the number of vertices in the graph.

\subsection{Signal reconstruction}
\label{subsec:reconstruction_methods}

Signal reconstruction is key to maintaining the data quality of a sensor network. Estimates can be obtained for a sensor that may present problems (e.g., missing value, malfunction, drift, etc.) using information from its neighboring sensors. In addition, in a worst case, such estimation can also be performed when the information is partial, i.e., there is more than one sensor whose signal needs to be reconstructed or presents missing data.
We consider the problem of having a subset of nodes (vertices of the graph) with samples and we would like to estimate the signal of the graph in the other vertices so that the resulting signal is smooth. This can be seen as a signal reconstruction problem that can be solved using methods from various fields. In this paper we consider the following; Laplacian interpolation, GSP low-pass based graph signal reconstruction, and kernel-based graph signal reconstruction\footnote{All Python code implementing the methods explained can be downloaded here at the research group \url{http://sans.ac.upc.edu/?q=node/231}.}. Laplacian interpolation \cite{belkin2004regularization} is a graph-based semi-supervised learning algorithm whose goal is regression with graph regularization assuming smoothness with respect to the Laplacian matrix. This method regresses a function $f{:}\mathcal{V}{\rightarrow}\mathbb{R}$ over the graph G, assuming partial information, it is to say, information for M nodes. GSP low-pass based graph signal reconstruction \cite{stankovic2019tutorial} is a graph signal processing reconstruction method that considers subsampling low-pass graph signals, thus assuming a sparse Fourier coefficient vector. Finally, kernel-based graph signal reconstruction \cite{romero2016kernel} formulates signal reconstruction as a regression task on reproducing kernel Hilbert spaces of graph signals. In the following, we briefly describe the three methods and their correspondence to a linear model.

\subsubsection{Laplacian interpolation} Also known as graph interpolated regularization by Belkin \textit{et al.} \cite{belkin2004regularization}, this method minimizes the quadratic form of the Laplacian matrix with respect to the graph signal $\bf{x}$, which is a measure of signal smoothness, given that the observed measurements $\{x_m{:\;} \forall m{\in}\mathcal{M}\}$ remain unchanged. This reconstruction results in a linear combination of the observations weighted by the Laplacian matrix entries $L_{ij}$.

\begin{equation}
\label{eq:smothness}
\begin{aligned}
& \underset{{\bf{y}}}{Min} & & {\bf{y}}^T \bf{L} \bf{y} \\
& \text{s.t.} & &  y_m = x_m,  \;\;\; \;\;\; \forall m\in \mathcal{M}, \\
\end{aligned}
\end{equation}


\subsubsection{Graph Signal Processing (GSP) low-pass reconstruction} This technique \cite{stankovic2019tutorial} recovers a set of unobserved nodes $\{x_u{:\;}\forall u{\in}\mathcal{U}\}$ given that the graph discrete Fourier transform of the complete signal is sparse and of low-pass nature, meaning that it has $K$ nonzero components corresponding to the lowest frequencies (smallest eigenvalues $\lambda_i$ of the Laplacian matrix). Given that the Laplacian matrix admits the eigendecomposition ${\bf{L}}{=} {\bf{U}}{\bf{\Lambda}}{\bf{U}}^T$, the graph discrete Fourier transform (GDFT) of a graph signal $\bf{x}$ can be computed as:
\begin{equation}
    {\bf{X}} = {\bf{U}}^{-1}{\bf{x}}
\end{equation}
Now, a K-sparse GDFT coefficient vector of the following form is to be recovered:
\begin{equation}
    {\bf{X}} = (X(0), \dots , X(K-1), 0, \dots , 0)^T
\end{equation}
For this purpose a subset of measurements $\mathcal{M}$ are used to recover the sparse coefficient vector by solving the following system:
\begin{equation}
    {\bf{x}}_{\mathcal{M}} = {\bf{U}}_{\mathcal{M}K}{\bf{X}}_K
\end{equation}
Since the system is overdetermined, the solution of the above system in the least squares sense is given by ${\bf{X}}_K {=\;} {\bf{U}}_{\mathcal{M}K}^\dagger{\bf{x}}_{\mathcal{M}}$, where ${\bf{U}}_{\mathcal{M}K}^\dagger{=\;}({\bf{U}}^T_{\mathcal{M}K}{\bf{U}}_{\mathcal{M}K})^{-1}{\bf{U}}^T_{\mathcal{M}K}$ is the matrix pseudo-inverse of ${\bf{U}}_{\mathcal{M}K}$; the nonzero coefficients are obtained, and after appending the corresponding zero coefficients, the inverse graph discrete Fourier transform (IGDFT) ${\bf{x}}{=}{\bf{U}}{\bf{X}}$ is computed to obtain the complete set of measurements ${\bf{x}}$ at all vertices.

\subsubsection{Kernel-based graph signal reconstruction} Romero \textit{et al.} \cite{romero2016kernel} introduced the kernelized reconstruction of graph signals. Given a set of noisy observations $\{x_m{=\;} y_0(v_m) + \epsilon_m{:\;} \forall m{\in}{\mathcal{M}}\}$, the kernel regression estimates the underlying function $y_0$ in a reproducing kernel Hilbert space (RKHS) $\mathcal{H}$, which is a space of functions $f{:}\mathcal{V}{\rightarrow}\mathbb{R}$:
\begin{equation}
    \mathcal{H}:=\{f{:} f(v)=\sum_{n=1}^{N} \alpha_n k(v, v_n), \alpha_n \in \mathbb{R}\}
\end{equation}
Where $k{:}\mathcal{V}{\times}\mathcal{V}{\rightarrow}\mathbb{R}$ is a kernel map that defines some similarity between nodes. After some manipulation and the application of the representer theorem, which states that the solution can be expressed as a linear combination of the kernel map values of the observed nodes $\hat{f_0}(v){=\;} \sum_{m \in \mathcal{M}} \hat{\bar{\alpha}}_m k(v, v_m)$, we can define the kernel ridge regression problem:
\begin{equation}
    \hat{\bar{\boldsymbol{\alpha}}} := \underset{\hat{\bar{\alpha}} \in \mathbb{R}^M}{\arg \min} \underbrace{\frac{1}{M}||{\bf{\bar{y}}} - {\bf{\bar{K}}}\boldsymbol{\bar{\alpha}}||^2}_{MSE} +  \mu \underbrace{\boldsymbol{\bar{\alpha}}^T{\bf{\bar{K}}}\boldsymbol{\bar{\alpha}}}_{RKHS \hspace{0.3em} norm}
\end{equation}
Where the mean squared error (MSE) is used as loss function, the regularization term is the RKHS norm of the solution $f_0$, $\boldsymbol{\bar{\alpha}}{=}\bf{\Phi}\boldsymbol{\alpha}$, ${\bf{\bar{K}}}{=}{\bf{\Phi}}{\bf{K}}{\bf{\Phi}}^T$ and $\bf{\Phi}$ is the sampling matrix. This problem has closed-form solution:
\begin{equation}
    \hat{\bar{\boldsymbol{\alpha}}} = ({\bf{\bar{K}}} + \mu M {\bf{I}}_M)^{-1}\bf{\bar{y}}
\end{equation}

The key to the successful application of kernel-based graph signal reconstruction is in the kernel selection. Thus, based on the assumption that the signal evolves smoothly over the graph, graph kernels that capture such prior information can be used, a common choice is the diffusion kernel (KRR-DIFF) where $r(\lambda){=\;} e^{\sigma^2\lambda/2}$:
\begin{equation}
    {\bf{K}} = r^\dagger({\bf{L}}) = {\bf{U}}r^\dagger ({\bf{\Lambda}}) {\bf{U}}^T
\end{equation}
Vertex-covariance kernel (KRR-COV) can also be used, which is based on the covariance instead of graph structure, which turns out to be the local linear minimum mean squared error (LMMSE) estimator on a Markov random field (MRF) \cite{romero2016kernel}. Since the actual covariance matrix is unknown, the graphical lasso has been used to estimate the covariance matrix $\bf{\hat{\Sigma}}$. Given the presence of multicollinearity in air pollution data \cite{ferrer2020graphsensing} and that the covariance matrix has to be estimated, the result is suboptimal.

\subsubsection{Linear models} \label{subsub:linear_models}
Given the analytical form of the solution of the methods described above, it can be noticed how the reconstruction of any of the methods corresponds to a linear combination ${\bf{x}}_{\mathcal{U}}{=\;}\boldsymbol{\beta}{\bf{x}}_{\mathcal{M}}$ of the observed nodes $\{x_m{:\;} \forall m{\in}\mathcal{M}\}$. Thus, Table \ref{tab:equivalence_linear} shows how the $\boldsymbol{\beta}$ coefficients are calculated for the reconstruction. As it can be seen, the main operations correspond to a matrix inversion and multiplication; in the case of Laplacian interpolation the $\boldsymbol{\beta}$ are calculated by the Laplacian $\bf{L}$, in the case of GSP by the Laplacian eigenvector matrix $\bf{U}$, in the case of the kernelized ridge regression with diffusion kernel by a pre-computed kernel matrix $\bf{K}$, and in the case of the kernelized ridge regression with vertex-covariance kernel by the data covariance matrix $\bf{\hat{\Sigma}}$.

\begin{table}[!htb]
\centering
\caption{Equivalence of the models to a linear model.}
\label{tab:equivalence_linear}
\resizebox{0.35\textwidth}{!}{%
\begin{tabular}{ll}
\hline \toprule
\textbf{MODEL:} & ${\bf{x}}_{\mathcal{U}}=\boldsymbol{\beta}{\bf{x}}_{\mathcal{M}}$ \\ \midrule
Lap.Int & $\boldsymbol{\beta}=-{\bf{L}}_{\mathcal{U} \mathcal{U}}^{-1}{\bf{L}}_{\mathcal{U} \mathcal{M}}$ \\
GSP & $\boldsymbol{\beta}={\bf{U}}_{\mathcal{U} K}({\bf{U}}_{\mathcal{M}K}^T{\bf{U}}_{\mathcal{M} K})^{-1}{\bf{U}}_{\mathcal{M}K}^T$ \\
KRR-DIFF & $\boldsymbol{\beta}={\bf{K}}_{\mathcal{U} \cdot}\boldsymbol{\Phi}^T(\boldsymbol{\Phi} {\bf{K}}\boldsymbol{\Phi}^T + \mu M {\bf{I}}_M)^{-1}$ \\
KRR-COV & $\boldsymbol{\beta}={\bf{\hat{\Sigma}}}_{\mathcal{U} \cdot}\boldsymbol{\Phi}^T(\boldsymbol{\Phi} {\bf{\hat{\Sigma}}}\boldsymbol{\Phi}^T + \mu M {\bf{I}}_M)^{-1}$  \\ \bottomrule \hline
\end{tabular}%
}
\end{table}


\begin{table*}[!htb]
\centering
\caption{Statistics of the data sets used for the signal reconstruction experiments.}
\label{tab:data_sets}
\resizebox{0.85\textwidth}{!}{%
\begin{tabular}{ccccccc}
\hline\toprule
\textbf{DATA SET} & \textbf{POLLUTANT} & \textbf{\# NODES} & \textbf{\# SAMPLES} & \textbf{PERIOD} & \textbf{MEAN} ($\mu gr/m^3$) & \textbf{POOLED STD.} ($\mu gr/m^3$) \\ \midrule
1 & O$_3$ & 46 & 1155 & 2019/01/01 - 2019/05/31 & 66.84 & 28.61 \\
2 & NO$_2$ & 60 & 983 & 2019/01/01 - 2019/05/31 & 23.46 & 16.52\\
3 & PM$_{10}$ & 33 & 709 & 2019/01/01 - 2019/05/31 & 20.11 & 11.51\\
4 & O$_3$ & 8 & 2612 & 2017/06/18 - 2017/09/16 & 64.92 & 34.84\\ \bottomrule \hline
\end{tabular}%
}
\end{table*}

\section{Data Sets}
\label{Sec:DS}
This section introduces the data sets used for the comparison of the reconstruction methods described above. Nowadays, there are many open data initiatives to improve transparency and encourage research. The Spanish government carries out the measurement of pollution levels by means of reference stations, which are worth thousands of euros given their high accuracy, and makes such data public. Therefore, data captured by reference stations\footnote{These data are available at \url{http://mediambient.gencat.cat/ca/05_ambits_dactuacio/atmosfera/qualitat_de_laire/vols-saber-que-respires/descarrega-de-dades/}.} in the area of Catalonia, Spain, over an area of 32,108 km$^2$ have been selected for three pollutants; tropospheric ozone (O$_3$), nitrogen dioxide (NO$_2$) and particulate matter 10 (PM$_{10}$).

These pollutants exhibit different spatial behavior, which allows studying the signal reconstruction under various spatial conditions. In addition, the use of reference stations over a large area makes it possible to investigate the feasibility of building a large network and the application of these methods. Table \ref{tab:data_sets} shows the characteristics of the data sets used, and Figure \ref{fig:maps_lap} presents the location of the reference stations for the data sets. Moreover, Table \ref{tab:data_sets} shows the means for the pollutants as well as the pooled standard deviation to better interpret the error measures in the following sections. These three data sets will be used to show the ability to reconstruct signals in an air pollution monitoring network in a regional area such as the metropolitan area of Barcelona, Spain.

The fourth data set corresponds to the deployment of a heterogeneous IoT network for the European project H2020 CAPTOR during the summer of 2017 \cite{barcelo2021h2020}. In this case, we set the data set to two reference stations and six nodes with low-cost sensors. These nodes are composed of Arduino as the processing unit, temperature, humidity, and five SGX Sensortech MICS 2614 metal oxide ozone sensors. The data are sent using a cellular network to a repository for off-line processing. This data set will be used to show a concrete application of signal reconstruction in a heterogeneous network when an sensor drifts.


\section{Results}
\label{Sec:Res}
The methodology used to learn the graph and train the signal reconstruction method is as follows; i) we use 100\% of the data for cross-validation (CV), ii) given the 2 hyperparameters of the graph learning method and the possible hyperparameters of the signal reconstruction method, we apply a 5-fold CV, iii) in each of the folds we learn the graph using the training data, we reconstruct each one of the nodes in the test and report the average root-mean squared error (RMSE) between all the nodes, and iv) given the hyperparameters corresponding to the lowest CV error we obtain the graph topology and the trained reconstruction method, section \ref{subsec:learning_graph}. From now on, all performance metrics reported throughout the paper are cross-validation metrics.

Three experiments are carried out for comparing the performance of the graph signal reconstruction methods. The first, section \ref{subsec:validation_error}, consists in reconstructing all nodes given that all neighbors are available during the CV, thus obtaining the average CV error to show the reconstruction power of the methods. The second, section \ref{subsec:semi_subsection}, simulates that a variable percentage of nodes need to be estimated simultaneously (e.g. some nodes may have missing data or are faulty nodes), this shows the ability of the methods to reconstruct the signals in a semi-supervised learning setting. The last experiment, section \ref{sub:scalability}, analyzes the scalability of the framework including both graph learning and signal reconstruction models. Finally, to conclude the results part, section \ref{sub:drift} shows a practical example of the signal reconstruction framework when the network compensates for sensor drift. The experiments in sections \ref{subsec:learning_graph} to \ref{sub:scalability} use data sets 1, 2 and 3, while the experiment in section \ref{sub:drift} uses data set 4.

\subsection{Learning an air pollution monitoring graph}
\label{subsec:learning_graph}
Figure \ref{fig:pipeline} shows the usual procedure for graph inference and signal reconstruction. First, given a data matrix $\bf{X}$ and the graph learning method with its hyperparameters $\alpha$ and $\beta$ (eq. \ref{eq:dong}), we get the Laplacian matrix $\bf{L}$ that describes the relationships between the nodes. Then, given a graph signal reconstruction model and its hyperparameters - none for the Laplacian interpolation, $K$ for the GSP, $\mu$ and $\sigma^2$ for KRR-DIFF and $\mu$ for KRR-COV - we can reconstruct a graph signal at any time $t$ given a subset of observed nodes $\mathcal{M}$. Actually, this is the methodology performed during the cross-validation, where the graph is learned using the training set of that fold and the nodes in the test set, of the same fold, are reconstructed. The four graph signal reconstruction models are transductive, it is to say, the coefficients for the reconstruction can be calculated given a set of observed nodes to reconstruct the unobserved ones, but if the set of observed nodes changes (e.g. some nodes have missings) the model needs to be recalculated to reconstruct the new set of unobserved nodes. Algorithm \ref{alg:gsr} gives a detailed description of the signal reconstruction framework process, where ${\bf{f}}(\cdot)$ is the graph signal reconstruction model, $C$ is the number of clusters, and $\{\boldsymbol{\alpha}, \boldsymbol{\beta}, \textbf{hyp}\}$ are the hyperparameters for the graph learning and graph signal reconstruction models. Lines 2 to 5 partition the set of nodes $\mathcal{V}$ into C disjoint clusters, where $\mathcal{C}_i{\subseteq}\mathcal{V}$ are the nodes indexes belonging to the $i$-th cluster, and learn a separate Laplacian ${\bf{L}}_i$ per cluster. We elaborate more on this idea in the scalability section \ref{sub:scalability}, for the moment we assume that we do not partition the set of nodes (i.e., C=1).


\begin{algorithm}
\small
\caption{Cluster-wise graph signal reconstruction.}
\label{alg:gsr}
\begin{algorithmic}[1]
    \INPUT $\{\boldsymbol{\alpha}, \boldsymbol{\beta}, {\bf{X}}, {\bf{f}}(\cdot ), \textbf{hyp}, C\}$
    \State $\bar{\bf{X}} \gets \text{Standardization}({\bf{X}})$
    \State $ \mathcal{C}_1,...,\mathcal{C}_C \gets \text{Clustering}(\bar{\bf{X}} ,C)$   \hfill\Comment{$\uparrow$ \textit{Scalability}}
    \FOR {$i=1 \; \text{to} \;C$}
          \State ${\bf{L}}_i \gets \text{Graph\_Learning}(\alpha_i, \beta_i, \bar{\bf{X}}_{\mathcal{C}_i})$
    \ENDFOR
    \WHILE {${\bf{x}}_{new}$} \hfill\Comment{\textit{New sample collected}}
        \State $\mathcal{U} \gets \text{Get\_Unobserved\_Nodes}({\bf{x}}_{new})$
	\State $\mathcal{M} \gets \mathcal{V}\setminus\mathcal{U}$
        \State $\bar{\bf{x}}_{new} \gets \text{Standardization}({\bf{x}}_{new})$
	 \FOR {$i=1 \; \text{to} \;C$}
           \State $\bar{{\bf{x}}}_{new_{{\mathcal{C}_{i_{\mathcal{U}}}}}} \gets {\bf{f}}(\bar{{\bf{x}}}_{new_{\mathcal{C}_{i_{\mathcal{M}}}}}, {\bf{L}}_i, \textbf{hyp}_i)$  \hspace{-30pt} \Comment{\textit{Reconstruction}}
    	\ENDFOR
        \State ${\bf{x}}_{new} \gets \text{Unstandardization}(\bar{{\bf{x}}}_{new})$
        \State $\text{return}\;\;{\bf{x}}_{new}$
    \ENDWHILE
\end{algorithmic}
\end{algorithm}

\begin{table*}
\centering
\caption{CV performance metrics for the data sets and reconstruction methods.}
\label{tab:CV_RMSE}
\resizebox{0.9\textwidth}{!}{
\begin{tabular}{cccccccccccccccc}
\hline \toprule
\multirow{2}{*}{\textbf{METHOD}} &  & \multicolumn{4}{c}{\textbf{O}$_3$} &  & \multicolumn{4}{c}{\textbf{NO}$_2$} &  & \multicolumn{4}{c}{\textbf{PM}$_{10}$} \\ \cline{3-6} \cline{8-11} \cline{13-16}
 &  & \textbf{RMSE} & \textbf{MAE} & \textbf{R}$^2$ & \textbf{\#EDGES} &  & \textbf{RMSE} & \textbf{MAE}  &\textbf{R}$^2$ & \textbf{\#EDGES} &  & \textbf{RMSE} & \textbf{MAE} & \textbf{R}$^2$ & \textbf{\#EDGES} \\ \midrule
\textbf{Lap.Int.} &  & 12.41 & 9.51 & 0.66 & 216.80 &  & 8.72 & 6.44 & 0.42 & 369.80 &  & 8.16 & 5.66 &  0.26 & 297.60 \\
\textbf{GSP} &  & 13.43 & 10.44& 0.56 & 1035.00 &  & 9.03 & 6.73 & 0.26 & 1761.20 &  & 8.54 & 5.99 &  0.16 & 528.00 \\
\textbf{KRR - DIFF} &  & 12.02 & 9.21 &  0.69 & 614.40 &  & 8.49 & 6.27 &  0.46 & 1514.00 &  & 8.10 & 5.61 &  0.29 & 481.80 \\
\textbf{KRR - COV} & & 11.72 & 8.98 & 0.71  & 328.80  &  & 8.23  & 6.04 &  0.50  & 638.40  &  & 8.03  & 5.51 &  0.30 & 198.00   \\ \bottomrule \hline
\end{tabular}
}
\end{table*}

\begin{figure}[!htb]
\centering
\subfigure[O$_3$ data set.]{\includegraphics[scale=0.43]{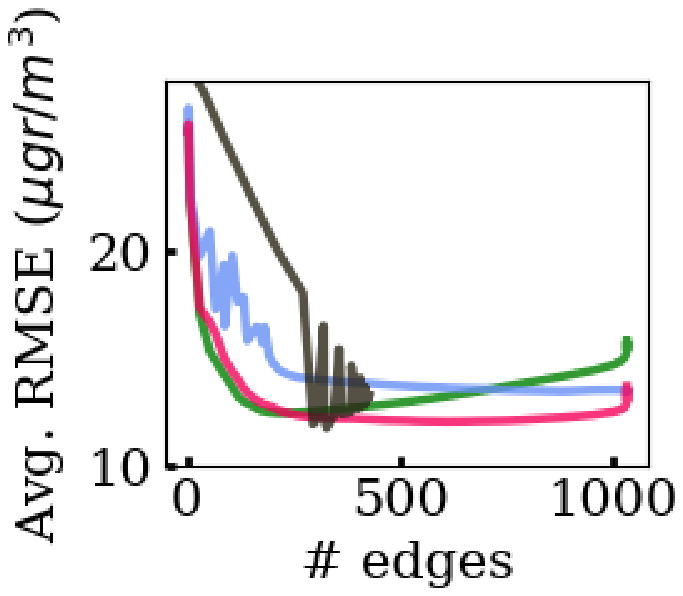}}
\subfigure[NO$_2$ data set.]{\includegraphics[scale=0.43]{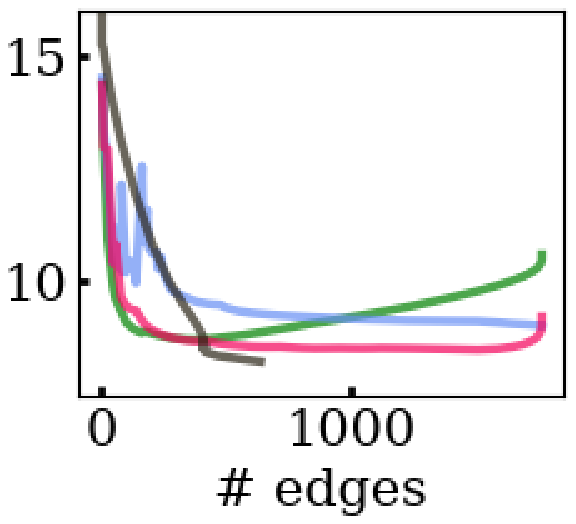}}
\subfigure[PM$_{10}$ data set.]{\includegraphics[scale=0.43]{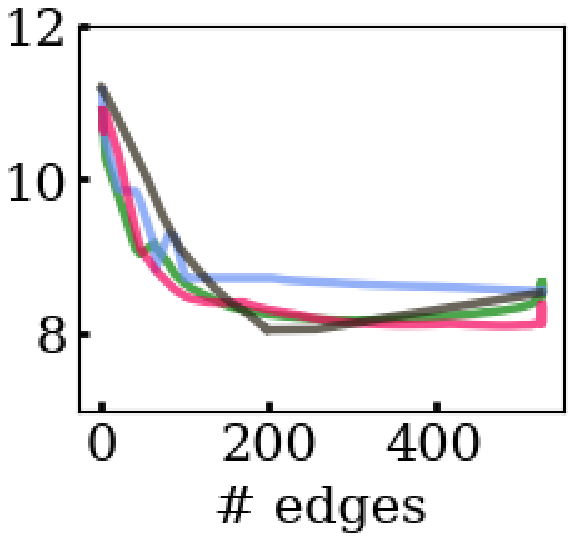}}
\caption{Average CV RMSE obtained for different number of edges and the best signal reconstruction hyperparameters. Line in green corresponds to the Lap.Int., blue line corresponds to GSP, and red and black lines correspond to KRR-DIFF and KRR-COV respectively.}
\label{fig:number_edges_rmse}
\end{figure}

There are then two blocks of hyperparameters; one that corresponds to graph learning and the other to signal reconstruction. Therefore, it is possible that for a given graph, with a certain number of edges, two methods perform in a different way. For this reason, cross-validation (CV) is made with all the hyperparameters, both for the graph learning and for the signal reconstruction, since the performance of the reconstruction method can be associated with the obtained graph. To illustrate this, Figure \ref{fig:number_edges_rmse} shows the average CV RMSE for each of the methods given the graph obtained. That is, the average RMSE of CV for each $\alpha$ and $\beta$ and the hyperparameters of the signal reconstruction method that have obtained the lowest error (the best $K$ for GSP, the best $\mu$ and $\sigma^2$ for KRR-DIFF and the best $\mu$ for KRR-COV). The case of the kernelized ridge regression with vertex-covariance kernel is a special case, where the covariance matrix is estimated using the graphical Lasso algorithm with different $\lambda$ hyperparameter values, thus obtaining an adjacency matrix (given the precision matrix $\boldsymbol{\Theta}{=}\boldsymbol{\Sigma}^{-1}$) but not a Laplacian matrix.

The same trends can be seen in the three plots in Figure \ref{fig:number_edges_rmse}. Laplacian interpolation gets its best error for a low number of edges, increasing later as the number of edges in the graph grows. The KRR-DIFF finds its lowest RMSE, similar to that of the Laplacian interpolation, but with a larger number of edges. The GSP reconstruction shows great instability for very sparse graphs, but its best error is obtained as the density of the graph increases. Finally, the case of KRR-COV shows great instability due to the multicollinearity present in the data, in many cases, it is not possible to obtain the complete graph with the graphical Lasso due to this problem. Also for similar graphs, the error can vary a lot, but the best error seems to be obtained with 50\% of the edges. So, for a low number of edges the Laplacian interpolation and KRR-DIFF are the best, the KRR-COV seems to obtain its best error with a relatively small number of edges, and the low-pass based reconstruction needs a large number of edges (around 100\%) to obtain its best performance. Although the KRR-COV is the best linear estimator for all three data sets, Figure \ref{fig:number_edges_rmse} shows how for sparse graphs ($<$25\% of the edges) the other methods produce a lower error. This is because the smoothness-based graph learning method is able to find a better set of neighbors per node for signal reconstruction than graphical Lasso, so although KRR-COV is optimal given a dense graph, it is not able to obtain the best performance for sparse graphs. In summary, each method behaves differently for a graph, so graph learning must be coupled to the signal reconstruction model in the cross-validation procedure.

\begin{figure*}[htb]
    \centering
    \includegraphics[width=0.93\textwidth]{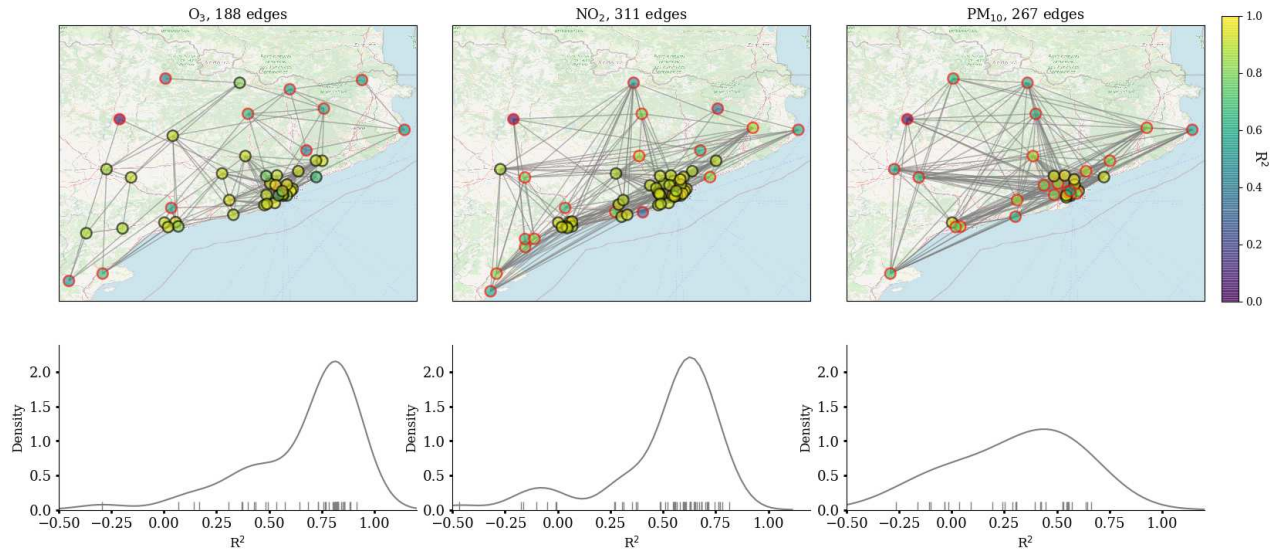}
    \caption{Plots representing the CV result for the data sets using the Laplacian interpolation method. Top figures show the graphs obtained with node color denoting the average cross-validation R$^2$ of each station. The nodes highlighted in red denote reference stations with a CV R$^2$ below 0.5. Bottom figures show the empirical R$^2$ distribution of the graphs above.}
    \label{fig:maps_lap}
\end{figure*}

\subsection{Signal reconstruction: All nodes available}
\label{subsec:validation_error}
Let us see the result of learning the graph and performing signal reconstruction given that the data of all node neighbors are available. 
The average CV error of all reference stations is used as performance metric. Table \ref{tab:CV_RMSE} shows the minimum average CV RMSE between stations, the average CV mean absolute error (MAE), as well as the corresponding average CV coefficient of determination R$^2$, and the average number of edges for the data sets and signal reconstruction methods. The methods appear to perform similarly, with KRR-COV being the best and GSP the worst. For the O$_3$ we can observe how the GSP is the worst method obtaining an R$^2$ of 0.56 with a complete graph (1035 edges) and the KRR with vertex-covariance kernel is the best model obtaining R$^2$ of 0.71 and 328.8 edges on average, followed by the KRR-DIFF that obtains a similar performance with a denser graph (614.40 edges on average) and the Laplacian interpolation whose best performance is obtained with a sparse graph. The same trend is observed for the NO$_2$ where the KRR-COV obtains the best performance with an R$^2$ of 0.50 and 638.40 edges on average and followed by the KRR-DIFF and Laplacian interpolation that obtain a denser and a sparser graph respectively. Finally, the reconstruction methods applied to the PM$_{10}$ data set obtain the worst results with R$^2$s around 0.16-0.3, but this time the KRR-COV performs the best with an average R$^2$ of 0.30 and the sparsest graph with 198.0 edges on average.


As it has been observed, O$_3$ and NO$_2$ can be estimated quite well by reconstructing the graph signal, but PM$_{10}$ can not. The reason is that PM$_{10}$ is more heterogeneous in the area of study than O$_3$ and NO$_2$, which makes it difficult to estimate using neighboring nodes. Among the methods, the KRR-COV obtains the best results as it can be interpreted as the local minimum linear mean squared error (LMMSE) estimator on a Markov random field. Nevertheless, as the covariance matrix is estimated and the data suffer from multicollinearity, the result is suboptimal and the graphical Lasso has not been able to obtain the covariance matrix with all dependencies (100\% of the edges). However, the KRR-DIFF performs almost as well as KRR-COV, followed by Laplacian interpolation which also shows good results.

The Laplacian interpolation has been able to obtain an error close to the optimal local linear estimator (KRR-COV) since the smoothness of the Laplacian matrix with respect to the training data is a criterion of the graph learning optimization problem. As for the resulting graphs, the GSP method obtains the best performance with 100\% of the edges. On the other hand, the Laplacian interpolation obtains its best performance with 21\% of the edges (with respect to the complete graph) for O$_3$, 21\% of the edges for NO$_2$ and 56\% of the edges for PM$_{10}$. This is an interesting result, since these methods allow obtaining a low RMSE with a small number of edges, which makes the graph sparse when the number of nodes is increased. Furthermore, in the previous section \ref{subsec:learning_graph}, it has been observed that the vertex-covariance method is no longer the best model for sparse graphs, with approximately less than 25\% of the edges.

Figure \ref{fig:maps_lap} shows the graphs obtained by the cross-validation procedure using the Laplacian interpolation method as reconstruction method. The color of the nodes denotes the average CV R$^2$ of the given stations. As shown in table \ref{tab:CV_RMSE}, despite having an average R$^2$ of 0.66, many of the stations in areas with a high density of stations (e.g. Barcelona area) have an R$^2$ around 0.8. Thus, ozone values in denser areas of similar reference stations can be estimated effectively. The same is observed for NO$_2$, although it presents a lower spatial correlation, some reference stations obtain a coefficient of determination larger than 0.7. Finally, PM$_{10}$ can not be approximated quite well using similar reference stations as most stations get an R$^2$ smaller than 0.6.

The results show a good prediction for most of the stations. Figure \ref{fig:maps_lap} shows the location of the nodes, some of them are far away from others or far from zones with density of stations. In the same figure, we can observe the reference stations highlighted in red, which are those with a CV R$^2$ smaller than 0.5. In the case of O$_3$, these are the most distant stations, and only 10 out of 46. In the case of NO$_2$, they already represent a larger percentage of the network stations with 19 stations out of 60 and are more distributed. Finally, in the case of PM$_{10}$ the bad reconstructed stations represent the majority, 22 stations out of 33. This represents the idea that O$_3$ and NO$_2$ are more predictable due to their homogeneity. This result indicates that if we look at the individual RMSE of the stations instead of the global one, those stations that are in a high-density area achieve a high R$^2$ and, therefore, a good estimate, while those that are more distant and, therefore, do not have many neighbors, do not take advantage of the nodes of the network as much. As an example, if we considered as outliers the stations with low R$^2$ and kept them out of the calculation, the average R$^2$ with the KRR-DIFF would improve from 0.69 to 0.78 for O$_3$ and from 0.46 to 0.65 in the case of NO$_2$. This shows the existence of a possible grouping of nodes into those that can be predicted well with their neighbors (group of similar nodes), and into other nodes that can not be predicted well (group of not similar nodes). This idea will be examined in more detail in the scalability section \ref{sub:scalability}, where similar nodes will be grouped using a clustering technique in order to improve the scalability.

\begin{figure*}[htb]
\centering
\subfigure[O$_3$ data set.]{\includegraphics[scale=0.45]{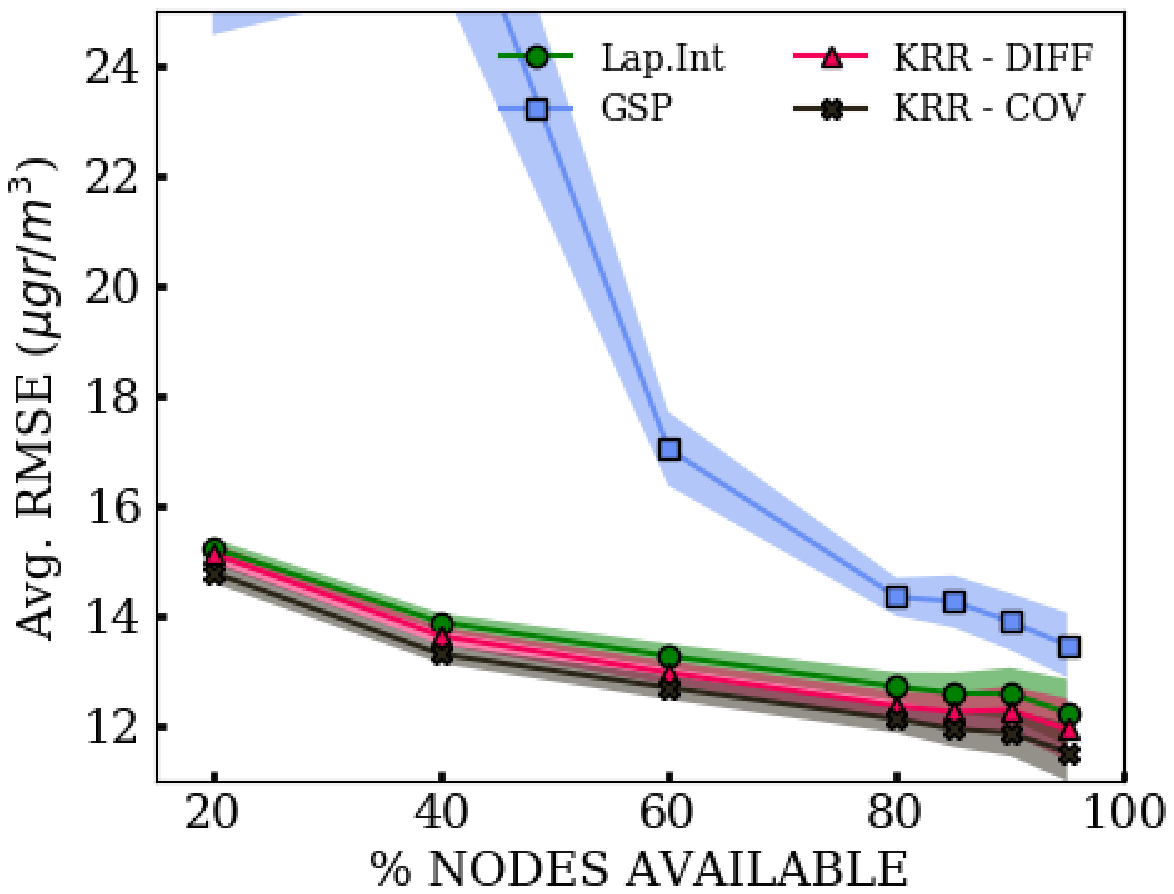}}
\subfigure[NO$_2$ data set.]{\includegraphics[scale=0.45]{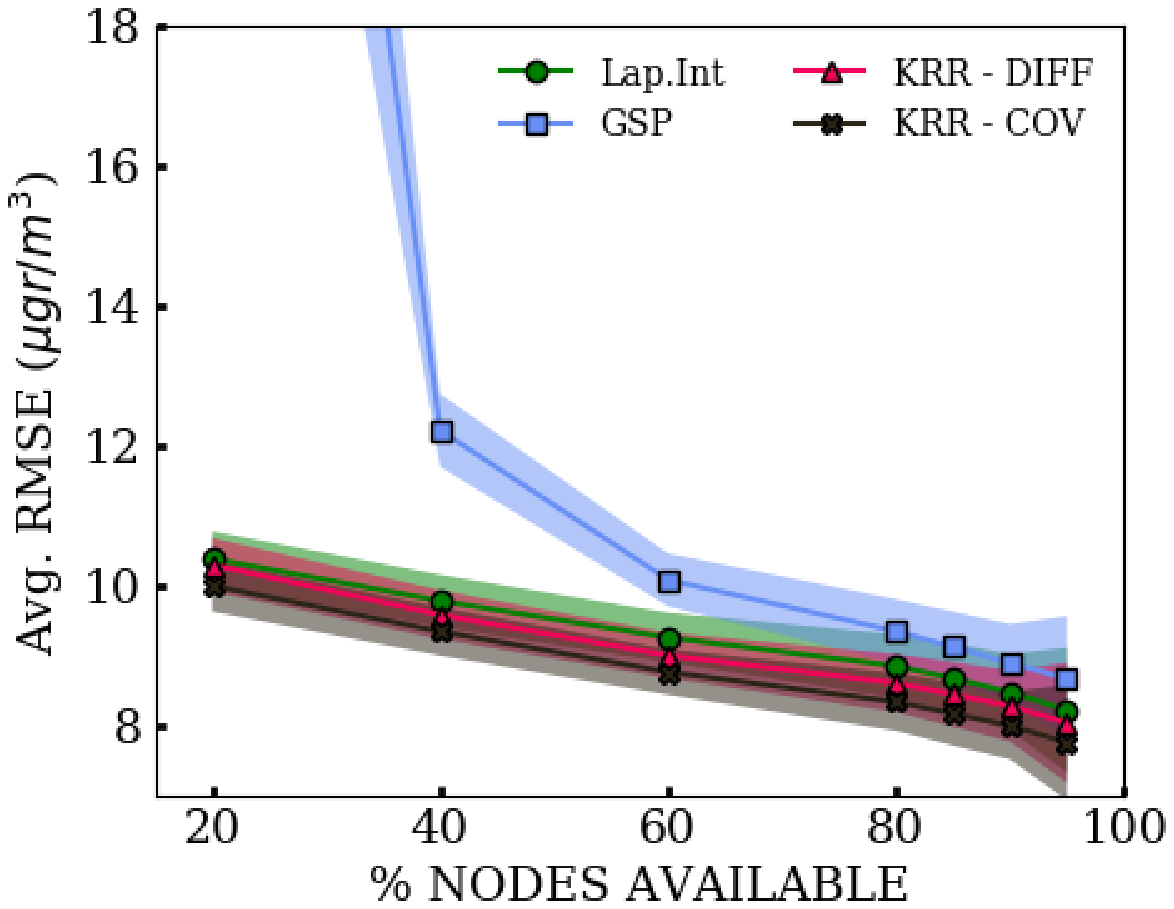}}
\subfigure[PM$_{10}$ data set.]{\includegraphics[scale=0.45]{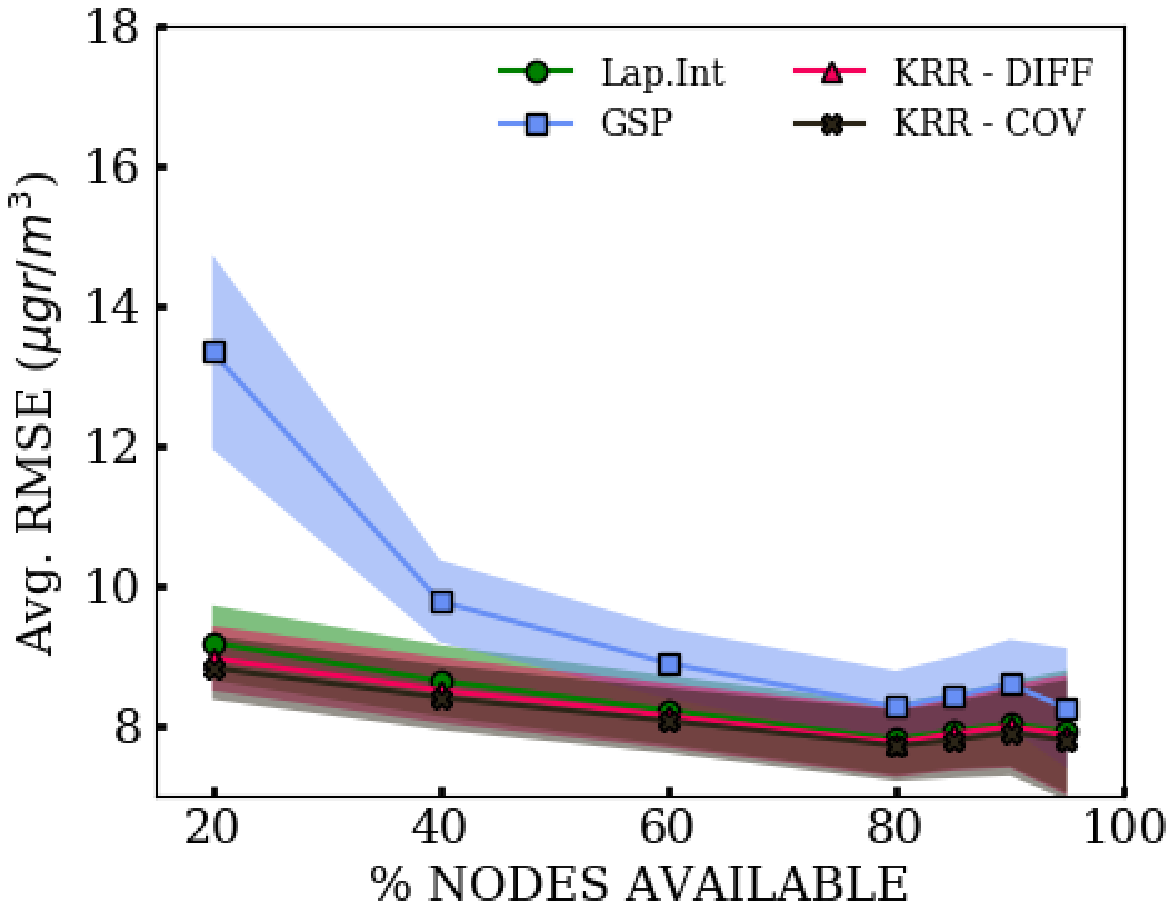}}
\caption{Average CV error and its 95\% confidence interval for several percentage of available nodes and 10 repetitions.}
\label{fig:semi_experiment}
\end{figure*}

\subsection{Signal reconstruction: Semi-supervised scenario}
\label{subsec:semi_subsection}
The previous section \ref{subsec:validation_error} showed the CV error of reconstructing the signal of a reference station when the others were available. In this section, we experiment with a semi-supervised learning setting, estimating multiple nodes at the same time, since we want to reconstruct several stations at once or we want to reconstruct some nodes that have missing data. This scenario is of special interest in the air pollution monitoring paradigm, since in a network there will always exist malfunctioning nodes, nodes with loss of samples, nodes in maintenance, and even the presence of virtual sensors. That is why the graph signal reconstruction model needs to be flexible to couple with the absence of data in neighboring sensors and estimate the signal in all nodes given a learned graph topology. Given a fixed set of hyperparameters, those found in the previous section, we calculate the CV error selecting a random incremental percentage of nodes to be estimated simultaneously, as if they were virtual sensors during the validation set, and perform ten repetitions.

Figure \ref{fig:semi_experiment} shows the mean CV error with bands indicating its 95\% confidence interval for the signal reconstruction methods and data sets. Figure \ref{fig:semi_experiment}.a) shows the results for the O$_3$ data set. First of all, we can observe how the GSP low-pass based reconstruction is the method that has the highest error with a 95\% of nodes available with 13.45 $\mu$gr/m$^3$ followed by a large difference by the Laplacian interpolation method with 12.22 $\mu$gr/m$^3$. In the previous section (100\% of nodes available), the Laplacian interpolation (12.41 $\mu$gr/m$^3$) already outperformed the GSP method (13.43 $\mu$gr/m$^3$). Moreover, as the percentage of available nodes decreases, the GSP low-pass based method is the reconstruction method whose error increases faster, rising from 60\% of nodes available. Thus, this method seems to be unable to generalize to the semi-supervised setting. The Laplacian interpolation and the kernelized ridge regressions are performing the best with a large difference. With 95\% of the nodes available, the kernelized ridge regression with a diffusion kernel obtains a CV RMSE of 11.91 $\mu$gr/m$^3$ and the KRR with the vertex-covariance kernel obtains a CV RMSE of 11.50 $\mu$gr/m$^3$, clearly the KRR-COV is the lower-bound of KRR-DIFF and Laplacian interpolation, and the three follow the same trend as the percentage of available nodes increases. In these latter cases, the error is slightly smaller than in the previous section probably due to the random selection of subsets of nodes, but it is noticed that with a high percentage of nodes available 95-80\% the error increases from 11.91 to 12.34 $\mu$gr/m$^3$ in the KRR-DIFF case, and from 11.86 to 12.27 $\mu$gr/m$^3$ in the vertex-covariance case. Finally, the error of all the reconstruction methods seems to grow significantly from 40\% of available nodes.

\begin{table*}
\centering
\caption{Methods' cost along with their hyperparameters.}
\label{tab:costs}
\resizebox{0.8\textwidth}{!}{%
\begin{tabular}{cccc} \hline \toprule
\textbf{METHOD} & \textbf{COST} & \textbf{HYPERPARAMETERS} & \textbf{OBSERVATIONS} \\ \midrule
\textbf{Graph learning} &\begin{tabular}[c]{@{}c@{}}Iterative algorithm, requires solving \\ iteratively a quadratic program and \\ a matrix $\mathbb{R}^{N\times N}$ inverse\end{tabular} & $\alpha$, $\beta$ & Easy to tackle cluster-wise \\
\textbf{Lap. Int} & \begin{tabular}[c]{@{}c@{}} Matrix inversion or multiplication \\ (depending on if $\left | \mathcal{U} \right |<<  \left | \mathcal{M} \right |$) \end{tabular} & None &  $\bf{L}_{\mathcal{U} \mathcal{U}}$ may be singular  \\
\textbf{GSP} &  \begin{tabular}[c]{@{}c@{}}  Matrix inversion or multiplication \\ (depending on if $K<<\left | \mathcal{M} \right |$) \end{tabular} & K &  \begin{tabular}[c]{@{}c@{}} Needs $\bf{L}=\bf{U}\bf{\Lambda}\bf{U}^T$ decomposition \\ once: $\mathcal{O}(N^3)$ \end{tabular} \\
\textbf{KRR-DIFF} & \begin{tabular}[c]{@{}c@{}} Matrix inversion or multiplication \\ (depending on if $\left | \mathcal{M} \right |<<\left | \mathcal{U} \right |$) \end{tabular} & $\mu$, $\sigma^2$ & \begin{tabular}[c]{@{}c@{}} Kernel matrix $\bf{K}$ only needs to be \\ computed once:  $\mathcal{O}(N^3)$ \end{tabular}\\
\textbf{KRR-COV} & \begin{tabular}[c]{@{}c@{}} Matrix inversion or multiplication \\ (depending on if $\left | \mathcal{M} \right |<<\left | \mathcal{U} \right |$) \end{tabular} & $\mu$ & \begin{tabular}[c]{@{}c@{}} Kernel matrix $\bf{K}$ only needs to be \\ computed once:  $\mathcal{O}(N^3)$ \end{tabular}\\ \bottomrule \hline
\end{tabular}%
}
\end{table*}

\begin{figure*}[htb]
    \centering
    \includegraphics[width=0.7\textwidth]{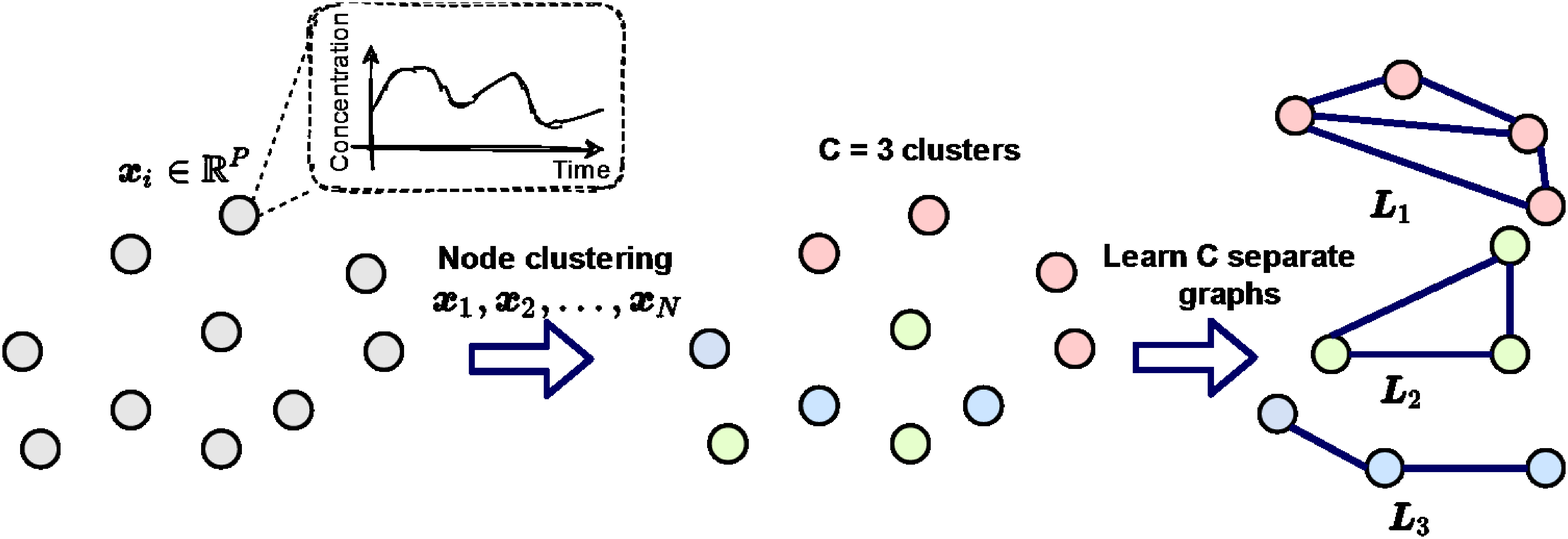}
    \caption{Example of air pollution monitoring graph learning process when dealing with large graphs, where C=3 clusters. The problem can be split and solved for C disjoint subgraphs.}
    \label{fig:cluster_process}
\end{figure*}

In Figure \ref{fig:semi_experiment}.b) the same results can be seen for the NO$_2$ data set. The GSP low-pass based method is the worst performing method, 8.66 $\mu$gr/m$^3$, with a high percentage of nodes available (95\%). The two KRR methods obtain similar errors, 8.03 $\mu$gr/m$^3$ and 7.75 $\mu$gr/m$^3$ respectively, in the case of NO$_2$, the difference between the two is even less than in the case of O$_3$. The kernel ridge regressions and the Laplacian interpolation methods show a similar pattern, the error with 95\% of the nodes (57 nodes) is less than the average error of all the nodes (subsection \ref{subsec:validation_error}) due to the random node subset selection. As the percentage of available nodes decreases the error increases, although the increase is moderate, the GSP low-pass method degrades considerably from 60\% nodes available, and in the kernel-based methods the increase is produced from 40 percent to 20 percent, going from 9.57 $\mu$gr/m$^3$ to 10.19 $\mu$gr/m$^3$ and 9.33 $\mu gr/m^3$ to 9.98 $\mu$gr/m$^3$ respectively. In summary, the kernelized ridge regressions and the Laplacian interpolation obtain similar values, the NO$_2$ has a smaller average concentrations (see section \ref{Sec:DS}) and these methods get lower average R$^2$ for the NO$_2$ data, since its concentrations are less smooth in an area than the O$_3$.

Finally, figure \ref{fig:semi_experiment}.c) shows the results for the particulate matter data set. Again, the methods follow the same trend, where as the percentage of available nodes increases the error decreases. It is worth remembering that in subsection \ref{subsec:validation_error} the PM$_{10}$ has been observed not to be predictable using neighboring concentrations, achieving a maximum coefficient of determination of 0.3 with the KRR-COV. Thus, the same results are maintained in the semi-supervised case where the worst method is GSP and the best the KRR-COV followed closely by its similar with diffusion kernel and the Laplacian interpolation. With a 20 percent of nodes available the methods get an error of; 13.32 $\mu$gr/m$^3$ with GSP, 9.17 $\mu$gr/m$^3$ with Laplacian interpolation, 8.95 $\mu$gr/m$^3$ with KRR-DIFF and 8.81 $\mu$gr/m$^3$ with KRR-COV.

In summary, we can conclude that Laplacian interpolation and kernel-based methods are robust and efficient in reconstructing the signal when an acceptable percentage ($>$60\%) of nodes are available, and that efficiency decreases as fewer nodes become available.

\subsection{Scalability}
\label{sub:scalability}
This section studies the scalability of the signal reconstruction framework, distinguishing between the graph learning task and the reconstruction of a graph signal $\bf{x}$ at a time instant $t$. Table \ref{tab:costs} shows the costs and the required hyperparameters for the methods. It is important to mention that the learning of the graph is linked to the training of the reconstruction method, since their hyperparameters are selected by means of a 5-fold CV. In this way, the more hyperparameters, the more dimensions the grid search will have and the more cost the cross-validation will have. 
 The graph learning task is solved by iteratively addressing a convex optimization problem that scales quadratically with the number of nodes $N$, so that, it is the most expensive task in the learning process. On the other hand, learning the graph for signal reconstruction with kernel ridge regression and diffusion kernel will require four hyperparameters ($\alpha$, $\beta$, $\mu$ and $\sigma^2$) and, therefore, will be the one that the grid search will take the longest. 

Once the hyperparameters are selected and thus the graph learned, it is not necessary to use a training set to train the graph signal reconstruction method, since these are transductive methods, where only the values of the set of observed nodes ${\bf{x}}_{\mathcal{M}}$ are needed to interpolate the unobserved ones ${\bf{x}}_{\mathcal{U}}$ at instant $t$. In the test set, the predictions will have the cost shown in the Table \ref{tab:costs} given the reconstruction models seen in section \ref{subsub:linear_models}. Generally, all four reconstruction methods require a matrix inversion or multiplication, whose cost is $\mathcal{O}(T^3)$, where T is the matrix dimension, but of different dimension. Thus, the cost comes from the dimension that dominates the operations (e.g. whether the number of unobserved nodes $\left | \mathcal{U} \right |$ is less than the observed ones $\left | \mathcal{M} \right |$). To speed up the reconstruction, if the set of nodes does not vary, the signal reconstruction coefficients can be calculated once and applied at different time instants $t$. However, if this set changes, which is a common situation where some of the nodes may have missing values and need to be reconstructed, the signal reconstruction coefficients must be recalculated. If cross-validation becomes extremely computationally expensive, a greedy approach can be carried out, where a graph with a desired level of sparsity is first obtained, and then cross-validation is performed only for the hyperparameters of the reconstruction method.

\begin{figure*}[!htb]
\centering
\subfigure[O$_3$ data set.]{\includegraphics[scale=0.43]{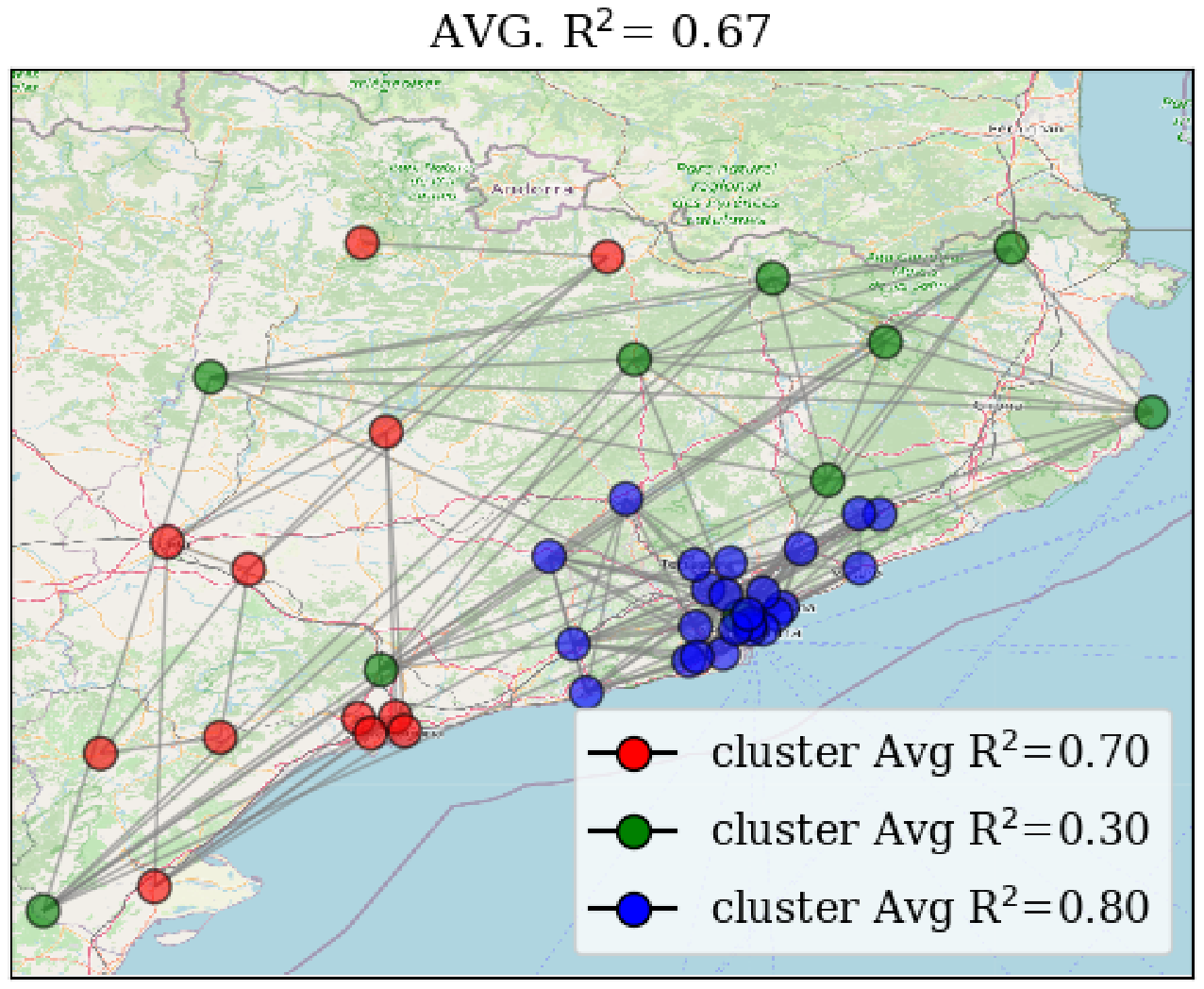}}
\subfigure[NO$_2$ data set.]{\includegraphics[scale=0.43]{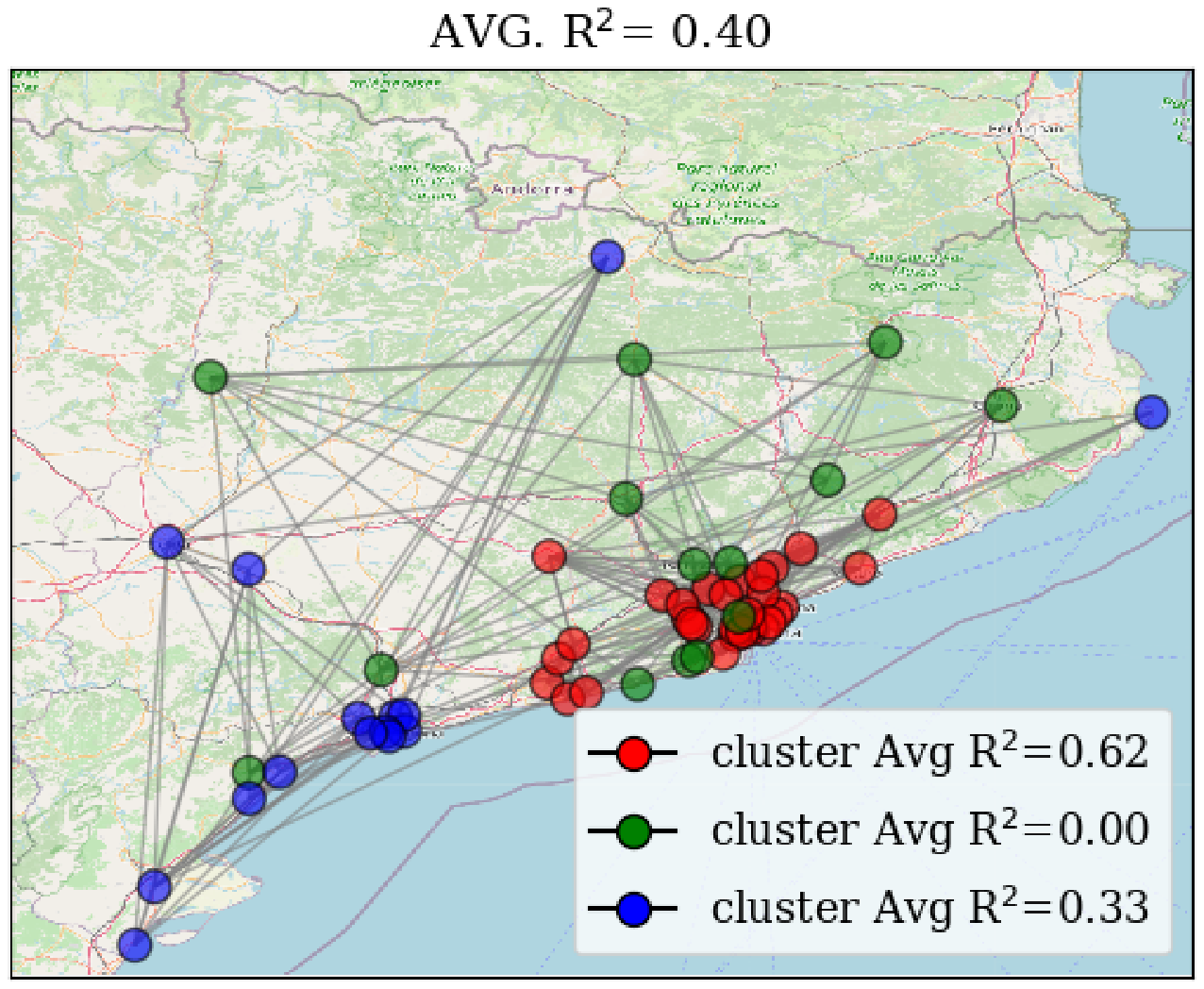}}
\subfigure[PM$_{10}$ data set.]{\includegraphics[scale=0.43]{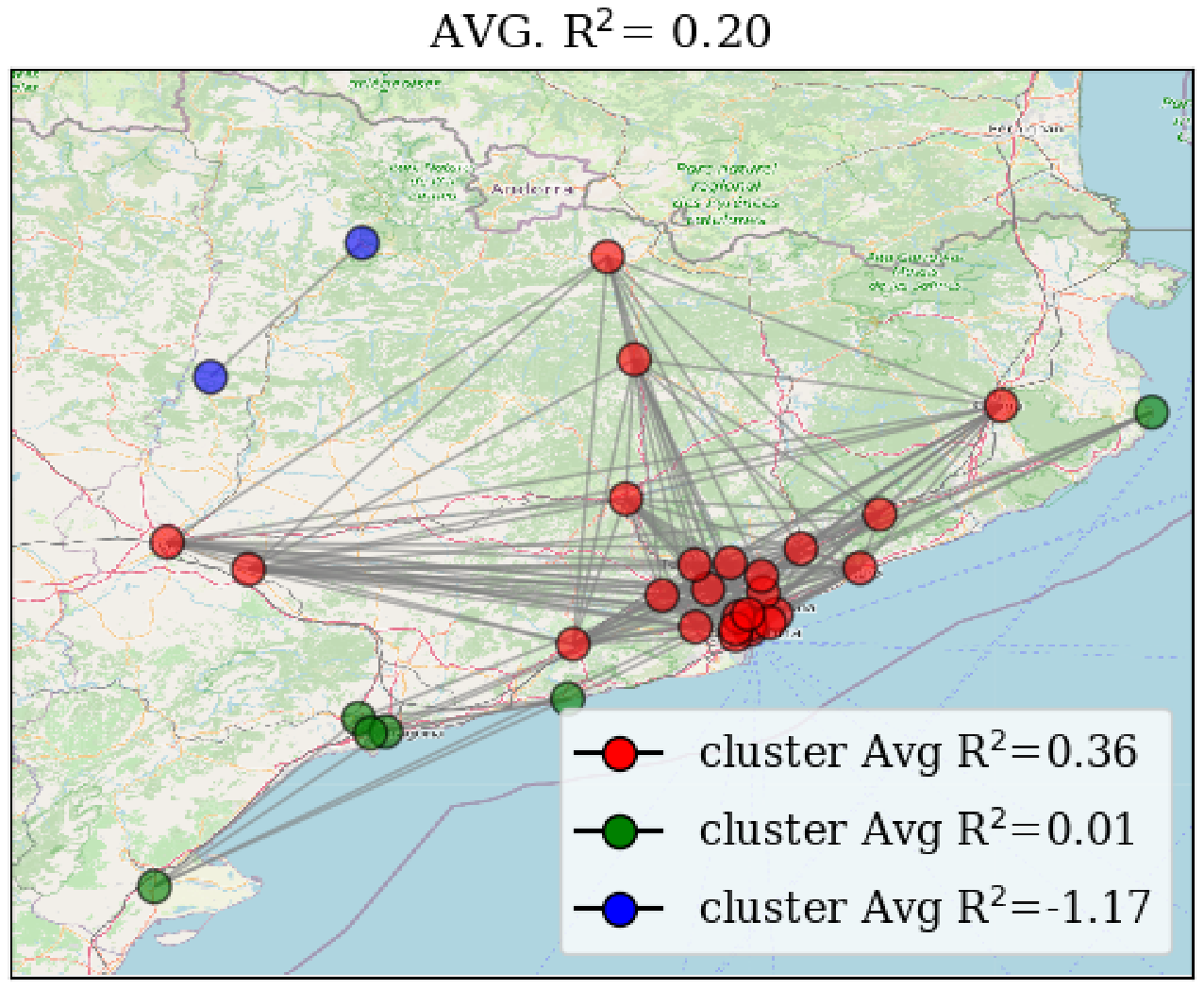}}
\caption{Results for a clustering example, with C=3, applied to all three data sets using the Laplacian interpolation.}
\label{fig:cluster_lap_int}
\end{figure*}

The most demanding task is graph learning since it involves iteratively solving a convex problem, and the complexity of this problem scales quadratically with the number of nodes N. However, as we are dealing with air pollution data, there exist spatial patterns and correlation between sensors, so the graph learning problem can be split to learn a number of disjoint subgraphs without having a large impact in the performance. Thus, we propose to find the approximation of a large graph as a set of disjoint subgraphs with a small impact on the signal reconstruction error. In a similar problem, Stein \textit{et al.} \cite{van2020cluster} propose a cluster-based methodology to solve the Kriging scalability issues, learning a single Kriging model per cluster. Therefore, given the correlations present in our data, a clustering-based approach to partitioning the graph learning problem is a good strategy to cope with the scalability problem.

Given the spatial correlation of certain pollutants (e.g. ozone tropospheric and nitrogen dioxide) and the assumption that stations with different variability patterns should be weakly connected, we divide the reference stations into C clusters and learn C Laplacians independently, resulting in C disjoint subgraphs. Figure \ref{fig:cluster_process} shows the graph learning process for large graphs, where using the normalized time series of each of the nodes $\bf{x}_i{\in}\mathbb{R}^P$, which can correspond to the training data, a clustering algorithm can be applied to the N nodes to find C clusters. The number of clusters can be decided in several ways, among them using unsupervised metrics such as the Calinski-Harabasz or the Silouhette index \cite{govender2020application}. Depending on the specific data set the size of the clusters may vary, but the computational complexity improvement of the problem will always depend on the size of the largest cluster, the smaller the greater the improvement, as the subgraph learning can be solved in parallel. The performance of the solution will always depend on the localization of the pollutants and the dependencies present in the data. In addition, the resulting Laplacians can be treated independently or joined together in a Laplacian in the form of a block-diagonal matrix. Algorithm \ref{alg:gsr} in section \ref{subsec:learning_graph} shows this cluster-wise procedure, where a different Laplacian ${\bf{L}}_i$ is learned for each cluster of nodes $\mathcal{C}_i$, with their corresponding $\alpha$ and $\beta$ hyperparameters. Once the Laplacians are learned, the graph signal reconstruction ${\bf{f}}(\cdot)$ can be applied in a cluster-wise manner, since nodes from different clusters are not connected.

\begin{table}[!ht]
\centering
\caption{Comparison of the average R$^2$ using the whole graph and using clusters.}
\label{tab:cluster_results}
\resizebox{0.85\columnwidth}{!}{%
\begin{tabular}{@{}ccccc@{}}
\toprule
\textbf{Data set} &  & \textbf{Original R$^2$} & \textbf{Cluster R$^2$} & \textbf{Problem size reduction}  \\  \midrule
O$_3$ & & 0.66 & 0.67 & $\downarrow 47.83\%$ \\
NO$_2$ & & 0.42 & 0.40 & $\downarrow 48.33\%$ \\
PM$_{10}$ & & 0.26 & 0.20 & $\downarrow 24.24\%$ \\\bottomrule
\end{tabular}%
}
\end{table}

Figure \ref{fig:cluster_lap_int}.a) shows the results for the first experiment (section \ref{subsec:validation_error}) learning three Laplacians (one per cluster) with the O$_3$ data set using the hierarchical clustering algorithm (colors denote the clusters obtained), which is a widely studied algorithm for clustering air pollution time series \cite{govender2020application}, using the euclidean distance and the ward linkage criterion. There is a cluster that has a very good average CV R$^2$ 0.80, another that has a good average R$^2$ with 0.7, and another set of stations that cannot be reconstructed well (R$^2$ of 0.30). Looking further into the results, we see how the cluster that is best predicted (blue cluster) is the one with the highest density of stations in an area and the second one (red cluster) has a few stations nearby. If we compare the results with those shown in Figure \ref{fig:maps_lap}, where we learned the graph with all the stations we see that the average CV R$^2$ is as good as the R$^2$ obtained learning the whole graph (0.66, see Table \ref{tab:CV_RMSE}). Moreover, it can be observed how in the clusters obtained for each pollutant the clustering algorithm is able to group significantly the nodes that do not benefit from the neighborhood information. In the case of NO$_2$ and PM$_{10}$, Figures \ref{fig:cluster_lap_int}b) and c) respectively, we see how the R$^2$ has worsened a little from 0.42 to 0.40 and from 0.26 to 0.20 with respect to learning the whole graph. In the case of PM$_{10}$, it is observed that in one of the clusters R$^2$=-1.17, with a negative value, which means that the model does not follow the trend of the data, and therefore fits worse than if there was a null model. These clusters correspond to nodes that cannot be reconstructed given their location and lack of influential neighboring information. Table \ref{tab:cluster_results} shows the average R$^2$ without using clustering and using clustering to improve the scalability. The coefficients of determination are mostly maintained or are slightly lower, so the scalability is improved at almost no signal reconstruction cost. The problem size reduction label indicates the reduction in the size of the resulting graph learning with respect to the original case, in the cluster-based learning case the size is defined by the size of the largest cluster. Thus, the problem size is reduced around a 24-48\%, meaning that the largest cluster is of size 75-52\% of the original graph, in almost all cases. As a summary, it is suggested that the problem of graph learning can be broken down into C problems with C clusters with the most similar reference stations, thus eliminating connections between subgraphs without a large impact on the error.


\subsection{Application: Drift compensation}
\label{sub:drift}

\begin{figure*}[!htb]
    \centering
    \subfigure[19-edge graph learned from the H2020 Captor data. ]{\includegraphics[scale=0.23]{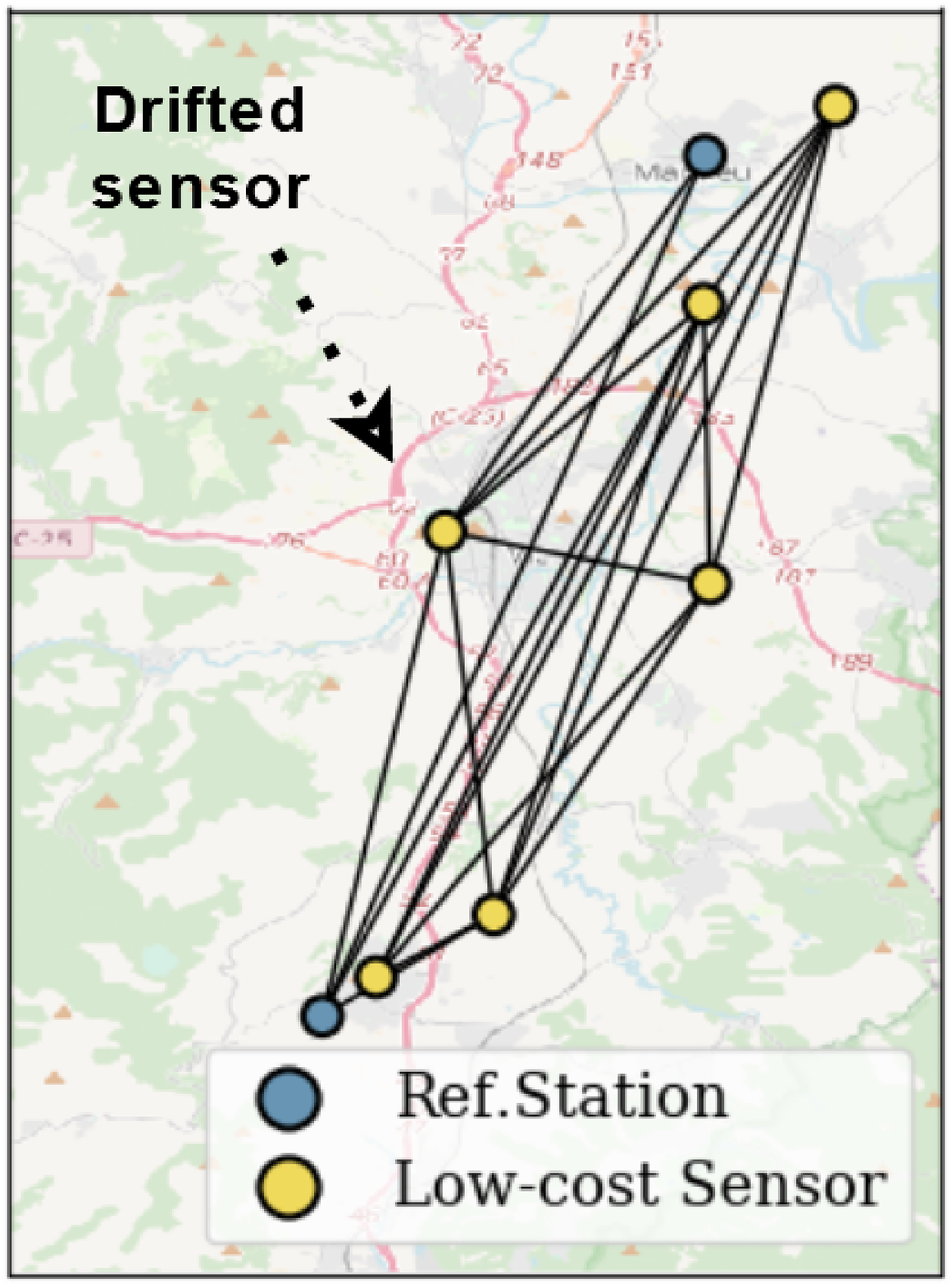}}
    \subfigure[Results for simulated drift compensation in two scenarios; using reference stations and low-cost sensors, and only using low-cost sensors.]{\includegraphics[width=0.56\textwidth]{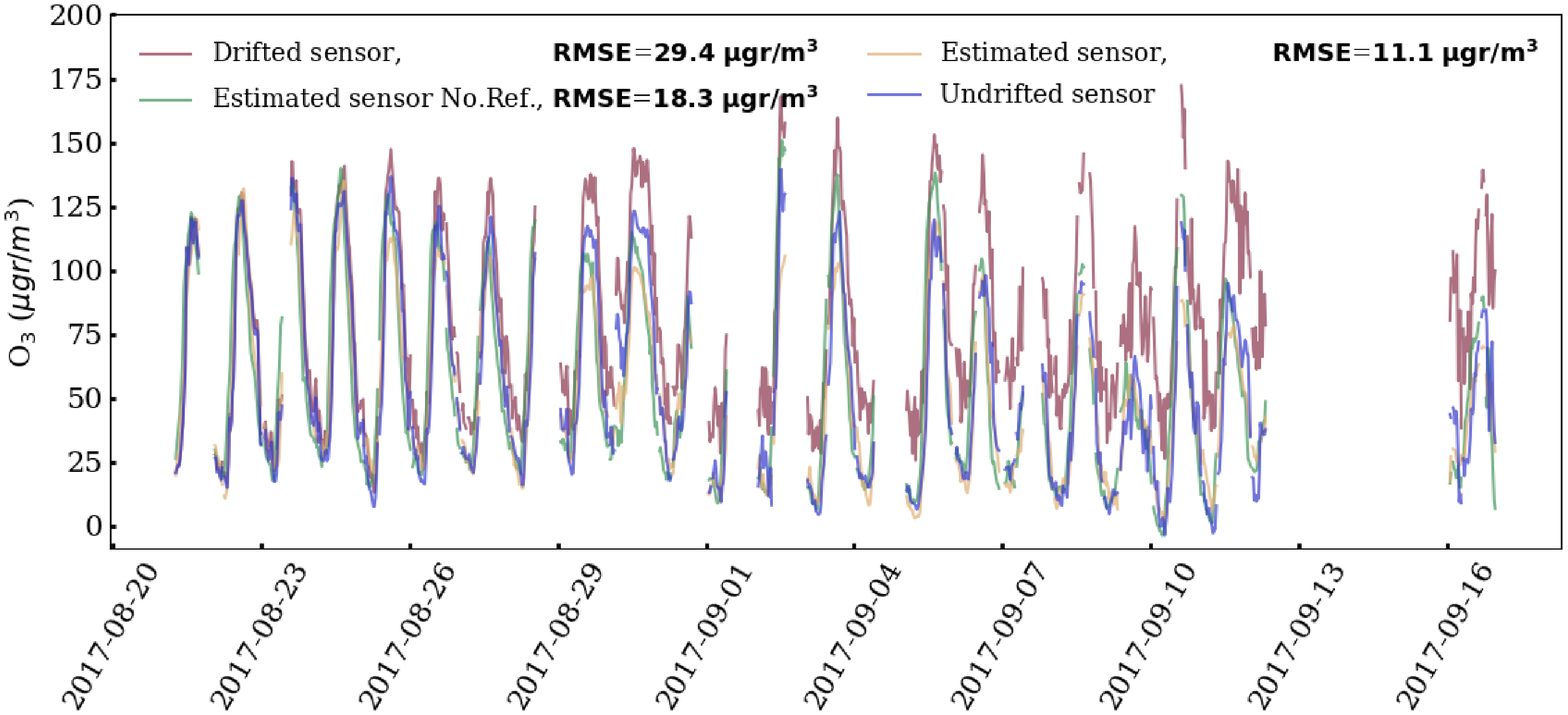}}
    \caption{Example of a simulated drifted sensor corrected with a heterogeneous sensor network using the signal reconstruction framework.}
    \label{fig:drift}
\end{figure*}

In this section, we show an example of signal reconstruction applied to a heterogeneous network, where a sensor drift is simulated by adding Gaussian noise of increasing magnitude. Signal reconstruction allows to maintain the quality of the drifting sensor by reconstructing its sensor values. We show how the compensation works when there are reference stations in the sensor neighborhood and when there are only other low-cost sensors. This small network containing eight nodes can be seen as a subgraph resulting from the split of a larger network using the clustering procedure shown in section \ref{sub:scalability}. Figure \ref{fig:drift}.a) shows the disposition of the nodes, the learned graph, the position of the drifted sensor and its six neighbors (two reference stations and four low-cost sensors). For illustration purposes, Laplacian interpolation has been used as a reconstruction method and the best graph has been learned performing cross-validation over the training set, 66\% of the data, and in the test set a drift has been simulated in the target sensor. Figure \ref{fig:drift}.b) shows the error made by the sensor with drift with respect to what would have been its normal behavior, 29.4 $\mu$gr/m$^3$. It is also observed the reconstruction made using all its neighbors, including the two reference stations, 11.1 $\mu$gr/m$^3$, showing the benefits from having precise instrumentation nearby to correct possible seasonal changes. Finally, as an interesting example, it has been simulated the unavailability of the two reference stations. Even in this case, the graph signal reconstruction is able to reduce the RMSE from 29.4 to 18.3 $\mu$gr/m$^3$ only using neighboring low-cost sensors, indicating the possibility of correcting drifts using only neighboring low-cost sensors. Alternatively, performing the same experiment using the kernelized ridge regression with vertex-covariance kernel and a 28-edge graph learned with the graphical lasso, the signal reconstruction error obtained using both reference stations and low-cost sensors is 10.89 $\mu$gr/m$^3$ and the error using only low-cost sensors is 14.84 $\mu$gr/m$^3$.

\section{Conclusions}
\label{Sec:Conc}

In this article, we have investigated the use of a signal reconstruction framework for air pollution monitoring data, where signal reconstruction techniques superimposed on a graph learned from the data can be used to maintain the data quality of this type of network. For this purpose, we have used O$_3$, NO$_2$ and PM$_{10}$ data from reference stations in Catalonia, Spain, and we have compared three reconstruction techniques from fields such as Laplacian interpolation (graph-based semi-supervised learning), low-pass graph signal reconstruction (signal processing) and kernel ridge regression with diffusion kernel and vertex-covariance kernel (kernel methods).

Firstly, the performance of each method has been studied by reconstructing each of the reference stations present in the network, assuming only missings in the node to be reconstructed. In this case, the kernel ridge regression methods and the Laplacian interpolation have demonstrated their superiority by outperforming the other method in the three data sets. Furthermore, it has been observed that O$_3$ and NO$_2$ can be correctly predicted with these methods as opposed to PM$_{10}$. The reason is the higher correlation between reference station data for O$_3$ and NO$_2$, while PM$_{10}$ data are more local and with lower correlation between stations. The kernel ridge regression with vertex-covariance kernel is the method that has worked best because it is the minimum local linear estimator, but the Laplacian interpolation and kernel ridge regression with diffusion kernel work better for sparse graphs. Secondly, we have studied the reconstruction with an increasing percentage of nodes to be estimated at the same time (e.g. nodes with missings) to simulate the semi-supervised learning paradigm. Again, the kernel methods and Laplacian interpolation outperform the other method and produce very good results when 0-40\% of the nodes are missing.

Thirdly, we have studied the scalability of proposed signal reconstruction framework, where the most expensive task is graph learning. Thus, since the cost of learning the graph involves a quadratic program that scales quadratically in the number of nodes in the network, the use of a clustering technique is proposed to approximate the problem of learning and reconstructing a large graph in the learning and reconstruction of C smaller graphs. Finally, the signal reconstruction framework has been successfully applied to a heterogeneous network to compensate for a drifting low-cost sensor. As future work, it would be interesting to study these ideas with other pollutants and the automatic application of signal reconstruction to correct faulty sensors in a sensor network deployment. In addition, other clustering algorithms or other partitioning techniques can be applied to allow high scalability of signal reconstruction methods on IoT platforms.

\bibliographystyle{IEEEtran}
\bibliography{references}


\begin{IEEEbiographynophoto}{Pau Ferrer-Cid} is a PhD student at the Statistical Analysis of Networks and Systems (SANS) research group, Universitat Politecnica de Catalunya (UPC). He holds a B.Sc in Computer Science and a M.Sc in Data Science by the UPC.  His main research interests are the applications of novel data analysis methods to sensor data coming from IoT platforms and the analysis of other kinds of data from fields like biology and computer vision.
\end{IEEEbiographynophoto}
\begin{IEEEbiographynophoto}{Jose M. Barcelo-Ordinas} is an Associate Professor at Universitat Politecnica de Catalunya (UPC) from 1999. He holds a PhD and B.Sc+M.Sc in Telecommunication Engineering and a B.Sc+M.Sc in Mathematics. He has participated in many European projects such as WIDENS, EuroNGI, EuroNFI, EuroNF NoE and H2020 CAPTOR. His currently research areas are wireless sensor networks, mobility patterns, and the statistical analysis of sensor data.
\end{IEEEbiographynophoto}
\begin{IEEEbiographynophoto}{Jorge Garcia-Vidal} is since 2003, full professor at the Computer Architecture Department of UPC, and since 2012 responsible of the Smart Cities Initiative at Barcelona Supercomputing Center (BSC-CNS), coordinating the H2020 CAPTOR project or being the BSC-CNS responsible of the H2020 project ASGARD. His main current research interest is in problems related with the capture, processing and statistical analysis of sensor data.
\end{IEEEbiographynophoto}

\end{document}